\documentclass[a4paper,12pt, leqno]{article}
\usepackage{amsmath}
\usepackage{amssymb, latexsym}
\usepackage{amsfonts}
\usepackage{makeidx}
\usepackage{amssymb}
\usepackage{color}     

\date{}

\newtheorem{lem}{Lemma}[section]
\newtheorem{thm}{Theorem}[section]
\newtheorem{prop}{Proposition}[section]
\newtheorem{cor}{Corollary}[section]

\numberwithin{equation}{section}
\newcommand{\dbar}{d\!\!\!{\lower-0.6ex\hbox{$-$}}\!}
\newcommand{\dslash}{d\!\!\!{\lower-0.6ex\hbox{$-$}}}
\newcommand{\e}{\varepsilon}
\newcommand{\h}{\hbar}
\newcommand{\ott}{\lower-0.4ex\hbox{${\scriptscriptstyle{\otimes}}$}}
\newcommand{\btt}{\lower-0.2ex\hbox{${\scriptscriptstyle{\bullet}}$}}
\newcommand{\ctt}{\lower-0.2ex\hbox{${\scriptscriptstyle{\circ}}$}}
\newcommand{\dtt}{\lower-0.2ex\hbox{${\scriptscriptstyle{\diamond}}$}}
\newcommand{\odt}{\lower-0.4ex\hbox{${\scriptscriptstyle{\odot}}$}}

\begin{document}

\title{Deformation Expression for Elements of Algebras (II)\\
--(Weyl algebra of $2m$-generators)--}


\author{
     Hideki Omori\thanks{ Department of Mathematics,
             Faculty of Sciences and Technology,
        Tokyo University of Science, 2641, Noda, Chiba, 278-8510, Japan,
         email: omori@ma.noda.tus.ac.jp}
        \\Tokyo University of Science
\and  Yoshiaki Maeda\thanks{Department of Mathematics,
                Faculty of Science and Technology,
                Keio University, 3-14-1, Hiyoshi, Yokohama,223-8522, Japan,
                email: maeda@math.keio.ac.jp}
          \\Keio University
\and  Naoya Miyazaki\thanks{ Department of Mathematics, Faculty of
Economics, Keio University,  4-1-1, Hiyoshi, Yokohama, 223-8521, Japan,
        email: miyazaki@hc.cc.keio.ac.jp}
        \\Keio University
\and  Akira Yoshioka \thanks{ Department of Mathematics,
          Faculty of Science, Tokyo University of Science,
         1-3, Kagurazaka, Tokyo, 102-8601, Japan,
         email: yoshioka@rs.kagu.tus.ac.jp}
           \\Tokyo University of Science
     }

\maketitle



\setcounter{equation}{0}

\tableofcontents


\bigskip 
This is a noncommutative version of the previous work \cite{OMMY3} 
in the same title with numbering $(I)$. In general in a noncommutative algebra,
there is no canonical way to express elements in univalent way, 
which is often called ``ordering problem''. In this note 
we discuss this problem in the case of the Weyl algebra of 
$2m$-generators. By fixing an expression, we extends Weyl algebra 
transcendentally. We treat $*$-exponential functions of linear 
forms, and quadratic forms of crossed symbol under generic expression  
parameters. 
In the extended algebra, we find a group which is isomorphic to 
$SO(m,{\mathbb C}){\times}{\mathbb Z}_2$ in the normal ordered 
expression, and several strange behaviour of $*$-exponential functions 
of quadratic forms.
After some investigation for $*$-exponential functions and intertwiners 
we find there is an expression parameter $K_s$ under which 
a certain system of $*$-exponential functions of quadratic forms 
generates a Clifford algebra. 
Hence, the transcendentally extended 
Weyl algebra has the property which may be called the 
Weyl-Clifford algebra.

\section{Weyl algebra in the normal ordered expression}

Throughout this paper, we use notations 
$$  
\tilde{\pmb u}=(\tilde{u}_1,\cdots,\tilde{u}_m),\,\,
\tilde{\pmb v}=(\tilde{v}_1,\cdots,\tilde{v}_m),\quad 
{\pmb u}=(u_1,u_2,\cdots,u_{2m})=
(\tilde{\pmb u},\tilde{\pmb v}).
$$
The {\bf Weyl algebra} $W_{\h}(2m)$ is the algebra generated by  
$\tilde{\pmb u}=(\tilde{u}_1,\cdots,\tilde{u}_m),\,\,
\tilde{\pmb v}=(\tilde{v}_1,\cdots,\tilde{v}_m)$ with the 
commutation relations
$$
[\tilde{u}_i,\tilde{v}_j]{=}-\h\delta_{ij},\quad 
[\tilde{u}_i,\tilde{u}_j]{=}0{=}[\tilde{v}_i,\tilde{v}_j], 
$$ 
where $[a,b]=a{*}b-b{*}a$ and $\h$ is a positive constant.
 When we consider the case $m=1$, 
$(\tilde{u},\tilde{v})$ stands 
for $(\tilde{u}_1,\tilde{v}_1)$ (or $(u_1,u_2)$). 
\medskip
One of the way for univalent expression of elements of $W_{\h}(2m)$ is 
to write $\tilde{\pmb u}$ in the l.h.s in each monomial. For instance,   
we write $\tilde{v}{*}\tilde{u}$, $\tilde{v}^2{*}\tilde{u}$
as $\tilde{u}{*}\tilde{v}{+}i\h$,
$\tilde{u}^2{*}\tilde{v}{+}2i\h\tilde{v}$. 
This way is called {\it normal ordering}.  

Furthermore, we identify a normally expressed element with a usual
polynomial and we denote, for instance 
$$
{:}\tilde{v}{*}\tilde{u}{*}\tilde{v}{+}\tilde{v}{*}\tilde{u}{:}_{_{K_0}}=
\tilde{u}\tilde{v}^2{+}\tilde{u}\tilde{v}{+}i\h\tilde{v}{+}i\h,\quad 
{:}\tilde{u}{:}_{_{K_0}}=\tilde{u},\quad
{:}\tilde{v}{:}_{_{K_0}}=\tilde{v},
$$ 
where the suffix ${\bullet}_{_{K_0}}$ indicate the normal ordered expression.
The sign ${*}$ is omitted in normally ordered expressions as  
they are ordinary polynomials.
 
For general $m$, the $*$-product (denoted by $*{_{_{K_0}}}$) of 
two normally ordered elements is given by the
abbreviate formula
\begin{equation}\label{psiDO00}
f({\pmb u}){*{_{_{K_0}}}}g({\pmb u})=
f\exp {\h i}\{\overleftarrow{\partial_{v}}\,\, 
       \overrightarrow{\partial_{u}}\}g,
       \qquad{\text{($\Psi$DO-product formula)}}
\end{equation} 
where  
$\overleftarrow{\partial_{v}}\,\, 
       \overrightarrow{\partial_{u}}
=\sum_i\overleftarrow{\partial_{\tilde{v}_i}}\,\, 
       \overrightarrow{\partial_{\tilde{u}_i}}$.
More precisely it is 
$$
f({\pmb u}){*{_{_{K_0}}}}g({\pmb u})=
\sum_k\frac{(i\h)^k}{k!}
\partial_{\tilde v_{i_1}}\cdots\partial_{\tilde v_{i_k}}f({\pmb u})
\partial_{\tilde u_{i_1}}\cdots\partial_{\tilde u_{i_k}}g({\pmb u}).
$$
We give several formulas which will be used later.
For $m=1$, we note first that the associativity and the commutation relation  
$\tilde u{*}\tilde v{-}\tilde v{*}\tilde u=-i\h$ gives for every polynomial 
$p(\tilde u{*}\tilde v)$ of $\tilde u{*}\tilde v$ that  
\begin{equation}\label{bump}
p(\tilde u{*}\tilde v){*}\tilde u=\tilde u{*}p(\tilde v{*}\tilde u)
=\tilde u{*}p(\tilde u{*}\tilde v{+}i\h),\quad (\text{bumping identity})
\end{equation}
Let $\tilde u{\ctt}\tilde v=
\frac{1}{2}(\tilde u{*}\tilde v{+}\tilde v{*}\tilde u)$; the 
symmetric product. The bumping identity gives  
$$
\tilde u{*}(\tilde u{*}\tilde v){*}\tilde v=
\tilde u{*}(\tilde u{\ctt}\tilde v{+}\frac{1}{2}i\h){*}\tilde v
=(\tilde u{\ctt}\tilde v
{+}\frac{1}{2}i\h){*}(\tilde u{\ctt}\tilde v{+}\frac{3}{2}i\h).
$$

\subsection{Star-exponential functions of crossed symbols}\label{crossed}

For every $C=(C_{ij})\in M_{\mathbb C}(m)$ ($m{\times}m$-complex
matrix), we denote 
$C(\tilde u,\tilde v)=\sum C_{ij}\tilde u_i\tilde v_j$. This special class of 
quadratic forms is very convenient for calculation in the normal ordered 
($K_0$-ordered) expression.  
In this section, normal ordered expressions are mainly used except
otherwise stated. 

\medskip  
By a direct calculation via $\Psi$DO-product formula \eqref{psiDO00} 
we have 
\begin{equation*}
  \begin{aligned}[t]
e^{\frac{2}{i\h}\sum A_{kl}\tilde u_k\tilde v_l}&{*}_{_{K_0}}
e^{\frac{2}{i\h}\sum B_{st}\tilde u_s\tilde v_t}=
e^{\frac{2}{i\h}\sum C_{ij}\tilde u_i\tilde v_j}, \\
C=&A+B+2AB.     
  \end{aligned}
\end{equation*}

If we denote $e^{\frac{2}{i\h}\sum A_{kl}\tilde u_k\tilde v_l}$ by
$[A]$, then the product formula is read as 
\begin{equation}
  \label{eq:noteigi}
[A]{*}_{_{K_0}}[B]=[A+B+2AB],\quad  A, B\in M_{\mathbb C}(m)   
\end{equation}
This is viewed as 
$$
(I+2A)(I+2B)=I+2(A+B+2AB).
$$
By the correspondence $A\leftrightarrow I+2A$, the multiplicative structure
of usual matrix algebra $M_{\mathbb C}(m)$ is translated 
into the space 
$\{e^{\frac{2}{i\hbar}C}; C\in M_{\mathbb C}(m)\}$. 
Let ${\mathcal O}'_{_{K_0}}=\{X\in M_{\mathbb C}(m); \det(I_m{+}2X)\not=0\}$. 

\begin{prop}\label{groupstr}
${\mathcal O}'_{_{K_0}}$ forms a group under the $*_{_{K_0}}$-product,  
 isomorphic to $G\!L(m,{\mathbb C})$. 
$0$ is the identity, and the inverse $X_*^{-1}$ is given by 
${-}(I{+}2X)^{-1}X$.
\end{prop}

But, note here that the additive unit $0\in M_{\mathbb C}(m)$ is 
shifted to $-\frac{1}{2}I_m(u,v)$, that is,  
$$
{-}\frac{1}{2}I{+}C{+}2({-}\frac{1}{2}I)C={-}\frac{1}{2}I, \quad i.e
$$
\begin{equation*}
e^{\frac{2}{i\hbar}C(\tilde u,\tilde v)}{*}_{_{K_0}}
e^{{-}\frac{1}{i\hbar}I_m(\tilde u,\tilde v)}=
e^{{-}\frac{1}{i\hbar}I_m(\tilde u,\tilde v)}{*}_{_{K_0}}
e^{\frac{2}{i\hbar}C(\tilde u,\tilde v)}=
e^{{-}\frac{1}{i\hbar}I_m(\tilde u,\tilde v)}.
\end{equation*} 
On the other hand, we see ${-}I{-}I{+}2(-I)(-I)=0$. In general 
\eqref{eq:noteigi} gives the following:
\begin{prop}\label{degenpolar} 
$e^{-\frac{2}{i\hbar}C(\tilde u,\tilde v)}{*}_{_{K_0}}
e^{-\frac{2}{i\hbar}C(\tilde u,\tilde v)}=e^{\frac{2}{i\h}0}=1$ 
if and only if $C^2=C$. In particular, 
let 
$$
I_k(\tilde u,\tilde v)=\tilde u_{i_1}\tilde v_{i_1}
{+}\tilde u_{i_2}\tilde v_{i_2}
{+}\cdots{+}\tilde u_{i_k}\tilde v_{i_k}
$$
for mutually distinct arbitrary $i_1,\cdots, i_k$. Then,
$e^{{-}\frac{2}{i\h}I_k(\tilde u,\tilde v)}{*}_{_{K_0}}
e^{{-}\frac{2}{i\h}I_k(\tilde u,\tilde v)}=1.$
\end{prop}

\medskip
By \eqref{eq:noteigi} we have the exponential law  
\begin{equation}\label{explaw1}
e^{\frac{1}{i\hbar}(e^{isC}{-}I_m)(\tilde u,\tilde v)}{*}_{_{K_0}}
e^{\frac{1}{i\hbar}(e^{itC}{-}I_m)(\tilde u,\tilde v)}=
e^{\frac{1}{i\hbar}(e^{i(s+t)C}{-}I_m)(\tilde u,\tilde v)}.
\end{equation}
Differentiate this exponential law \eqref{explaw1} to obtain  
the $*_{_{K_0}}$-exponential function  
\begin{equation}
 \label{eq:norexpm}
{:}e_*^{\frac{it}{i\hbar}\sum C_{kl}\tilde u_k{*}\tilde v_l}{:}_{_{K_0}}=
e^{\frac{1}{i\hbar}\sum(e^{itC}{-}I_m)_{kl}\tilde u_k\tilde v_l} 
\end{equation}
where $\tilde u_k{*}\tilde v_k$ stands for ${:}\tilde u_k{*}\tilde v_k{:}_{_{K_0}}$, but 
this is $\tilde u_k\tilde v_k$ in the normal ordered expression. 
Here, we used the notation such as 
${:}e_{*}^{\frac{it}{i\hbar}\sum C_{kl}\tilde u_k{*}\tilde v_l}{:}_{_{K_0}}$
to avoid possible confusion. 
It is remarkable that the amplitudes are kept to be $1$ in this calculation.

\medskip
In what follows of this section, we use $\tilde u_i{\ctt}\tilde v_j$ often instead
 of $\tilde u_i{*}\tilde v_j$, where 
$$
\tilde u_i{\ctt}\tilde v_j=\frac{1}{2}(\tilde u_i{*}\tilde v_j{+}\tilde v_j{*}\tilde u_i),\quad 
{:}\tilde u_i{\ctt}\tilde v_j{:}_{_{K_0}}=\tilde u_i\tilde v_j{+}\frac{1}{2}i\h \delta_{ij}.
$$
The reason of this is that the expression by using 
$\tilde u_i{\ctt}\tilde v_j$ has a rich symmetric properties, just
like Stratonovich formula in stochastic integrals.

\medskip
By using this notation, \eqref{eq:norexpm} is changed into   
\begin{equation}\label{eq:norexpm2}
{:}e_{*}^{\frac{t}{i\hbar}\sum C_{kl}\tilde u_k{\ctt}\tilde v_l}{:}_{_{K_0}}=
e^{\frac{t}{2}{\rm{Tr}}C}
e^{\frac{1}{i\hbar}\sum(e^{tC}{-}I)_{kl}\tilde u_k\tilde v_l}
\end{equation}
with a nontrivial amplitude part, but note here that 
the amplitude part is determined by its phase part.
If $C^2=C$, then $e^{tC}=I{+}(e^t{-}1)C$, hence 
\begin{equation}\label{polar0101}
{:}e_{*}^{\frac{\pi i}{i\hbar}\sum C_{kl}\tilde u_k{\ctt}\tilde v_l}{:}_{_{K_0}}=
e^{\frac{\pi i}{2}{\rm{Tr}}C}
e^{-\frac{2}{i\hbar}\sum C_{kl}\tilde u_k\tilde v_l}, \quad 
{:}e_{*}^{-\frac{\pi i}{i\hbar}\sum C_{kl}\tilde u_k{\ctt}\tilde v_l}{:}_{_{K_0}}=
e^{-\frac{\pi i}{2}{\rm{Tr}}C}
e^{-\frac{2}{i\hbar}\sum C_{kl}\tilde u_k\tilde v_l}.
\end{equation}
A typical example is 
$$
C(\theta)=
\begin{bmatrix}
\cos^2\theta&\cos\theta\sin\theta\\
\cos\theta\sin\theta&\sin^2\theta\\
\end{bmatrix}
,\quad 
{:}e_{*}^{\pm\frac{\pi i}{i\hbar}\sum C_{kl}(\theta)\tilde u_k{\ctt}\tilde v_l}{:}_{_{K_0}}=
(\pm i)e^{-\frac{2}{i\hbar}\sum C_{kl}(\theta)\tilde u_k\tilde v_l}.
$$

\bigskip
Consider now the group ${\mathfrak G}$ generated by 
${:}e^{\frac{2}{i\h}(\sum A_{kl}\tilde u_k{\ctt}\tilde v_l)}{:}_{_{K_0}}$,  
$A\in{\mathcal O}'_{_{K_0}}$ and two homomorphisms
$$
\pi: {\mathfrak G}\to [{\mathcal O}'_{_{K_0}}],\quad \tilde{\alpha} : {\mathfrak G}\to{\mathbb C}_*
$$
defined by 
$$
\pi({:}e^{\frac{1}{i\h}(\sum A_{kl}\tilde u_k{\ctt}\tilde v_l)}{:}_{_{K_0}})=
e^{\frac{1}{i\h}(\sum A_{kl}\tilde u_k\tilde v_l)}, \quad 
\tilde{\alpha}({:}e^{\frac{1}{i\h}(\sum A_{kl}\tilde u_k{\ctt}\tilde v_l)}{:}_{_{K_0}})=e^{\frac{1}{2}{\rm{Tr}}A}.
$$
As it is shown in  Proposition\,\ref{degenpolar}, and in \eqref{eq:norexpm2} even though
$\pi(X){=}1$, it may occur $\tilde\alpha(X){=}{-}1$.

\begin{prop}\label{forgettingamp}
In the normal ordered expression, 
${:}e^{\frac{2}{i\h}(\sum A_{kl}\tilde u_k{\ctt}\tilde v_l)}{:}_{_{K_0}}$,
$A\in{\mathcal O}'_{_{K_0}}$ generates 
a group isomorphic to a connected double covering of 
$G\!L(m,{\mathbb C})$ embedded naturally in  
${\mathbb C}_*{\times}G\!L(m,{\mathbb C})$ in the form 
$(\tilde{\alpha}(X),\pi(X)), X\in {\mathfrak G}$.
\end{prop}

Our main concern in this section is elements 
$$
\e_{00}(k) =e_{*}^{\frac{\pi i}{i\hbar}\tilde u_k{\ctt}\tilde v_k}=
 e_*^{\frac{\pi}{2\hbar}(\tilde u_k*\tilde v_k+\tilde v_k*\tilde u_k)}, \quad 
k=1, 2, \dots, m.
$$
Although, $e_{*}^{\pm\frac{\pi i}{i\hbar}\tilde u_k{\ctt}\tilde v_k}$ are defined 
only for the normal ordered expression at this stage, we denote it by  
\begin{equation}\label{polarpolar}
{:}{\e}_{00}(k){:}_{_{K_0}} = {:}e_{*}^{\frac{\pi i}{i\hbar}\tilde u_k{\ctt}\tilde v_k}{:}_{_{K_0}}{=}
{:}e_{*}^{\frac{\pi i}{i\hbar}(\tilde u_k{*}\tilde v_k{+}\frac{i\h}{2})}{:}_{_{K_0}}{=}
{:}{i}e_{*}^{\frac{\pi i}{i\hbar}\tilde u_k{*}\tilde v_k}{:}_{_{K_0}}{=}
ie^{-\frac{2}{i\h}\tilde u_k\tilde v_k},
\end{equation}
for we will define in the later section an abstract 
$e_{*}^{\frac{\pi i}{i\hbar}\tilde u_k{\ctt}\tilde v_k}$ in generic ordered expressions.  
We call ${\e}_{00}(k)$ {\it partial polar elements} and
${:}{\e}_{00}(k){:}_{_{K_0}}$ its {\it normal ordered expression}.  We see  
$$
{:}\e_{00}(k){:}_{_{K_0}}=ie^{-\frac{2}{i\h}\tilde u_k\tilde v_k},\quad 
{:}\e_{00}(k){*}\e_{00}(k){:}_{_{K_0}}=-1
$$
${}$\hfill (In the later section, we see Weyl ordered expression of $\e_{00}(k)$ diverges.)

By the product formula under the normal ordered expression $K_0$, the natural commutativity  
\begin{equation}\label{commute}
{:}e_{*}^{\frac{\pi i}{i\hbar}\tilde u_k{*}\tilde v_k}{:}_{_{K_0}}
{*_{_{K_0}}}
{:}e_{*}^{\frac{\pi i}{i\hbar}\tilde u_l{*}\tilde v_l}{:}_{_{K_0}}{=}
{:}e_{*}^{\frac{\pi i}{i\hbar}(\tilde u_k{*}\tilde v_k{+}\tilde u_l{*}\tilde v_l)}{:}_{_{K_0}}
{=}{:}e_{*}^{\frac{\pi i}{i\hbar}\tilde u_l{*}\tilde v_l}{:}_{_{K_0}}
{*_{_{K_0}}}{:}e_{*}^{\frac{\pi i}{i\hbar}\tilde u_k{*}\tilde v_k}{:}_{_{K_0}}
\end{equation}
holds. Summarizing these we have 
\begin{prop}\label{polar}
${:}\e_{00}(k){*}\e_{00}(\ell){:}_{_{K_0}}={:}\e_{00}(\ell){*}\e_{00}(k){:}_{_{K_0}}$,\quad 
${:}\e_{00}(k){*}\e_{00}(k){:}_{_{K_0}}=-1$
\end{prop}

However, {\bf this does not imply that  ${\e}_{00}(k)$ 
are commuting each other in general nor} ${\e}_{00}(k)_*^2=-1$ {\bf in general}. 
Indeed, it will be shown in \S\,\ref{CiffCliff} that there is 
an expression parameter $K_s$ such that ${\e}_{00}(k)$ are 
mutually anti-commuting under $K_s$-expression. Moreover, in
\S\,\ref{Sieg} we see there is a class of expression parameters $K$
such that  ${:}{\e}_{00}(k)_*^2{:}_{_K}=1$.

We set 
${\e}_{00}(I)$  by $e_{*}^{\frac{\pi i}{i\hbar}\sum_k\tilde u_k{*}\tilde v_k}$ 
and call it the {\bf total polar element}.  
Its $K_0$-expression is 
\begin{equation}\label{polarpolar00}
{:}{\e}_{00}(I){:}_{_{K_0}}={i^m}e^{-\frac{2}{i\h}\sum_{k=1}^m \tilde u_k\tilde v_k}.
\end{equation}

\bigskip
If $\langle\tilde{\pmb a}, \tilde{\pmb a}\rangle=\tilde{\pmb a}{}^{t}\!{\tilde{\pmb a}}=1$, then by noting 
$({}^{t}\!\tilde{\pmb a}\tilde{\pmb a})^n
={}^{t}\!\tilde{\pmb a}\tilde{\pmb a}$, we see 
$$
e^{s{}^{t}\!\tilde{\pmb a}\tilde{\pmb a}}
=I+(e^s-1){}^{t}\!\tilde{\pmb a}\tilde{\pmb a}
$$

For $\tilde{\pmb a}, \tilde{\pmb b}\in{\mathbb C}^m$, we set 
$\langle\tilde{\pmb a},\tilde{\pmb u}\rangle=\sum_{i=1}^m \tilde a_i\tilde u_i$,
$\langle\tilde{\pmb a},\tilde{\pmb v}\rangle=\sum_{i=1}^m \tilde a_i\tilde v_i$ etc. 

Then, it is easy to see 
$[\langle\tilde{\pmb a},\tilde{\pmb u}\rangle, 
  \langle\tilde{\pmb b},\tilde{\pmb v}\rangle]=-\hbar i\langle\tilde{\pmb a},\tilde{\pmb b}\rangle$.
Hence, if 
$\langle\tilde{\pmb a},\tilde{\pmb a}\rangle=1$, then 
$\langle\tilde{\pmb a},\tilde{\pmb u}\rangle$ and 
$\langle\tilde{\pmb a},\tilde{\pmb v}\rangle$ form a canonical conjugate pair.

Let
${S}_{\mathbb C}^{m-1}{=}\{{\pmb a}\in 
{\mathbb C}^m; \langle\tilde{\pmb a},\tilde{\pmb a}\rangle{=}1\}$, 
and   
${S}_{\mathbb R}^{m-1}{=}\{\tilde{\pmb a}\in 
{\mathbb R}^m; \langle\tilde{\pmb a},\tilde{\pmb a}\rangle{=}1\}$.
Then, for every $\tilde{\pmb a}\in {S}_{\mathbb C}^{m-1}$, the quadratic form 
$$
\alpha\langle\tilde{\pmb a},\tilde{\pmb u}\rangle_*^2+\beta\langle\tilde{\pmb a},\tilde{\pmb v}\rangle_*^2+
2\gamma\langle\tilde{\pmb a},\tilde{\pmb u}){\ctt}\langle\tilde{\pmb a},\tilde{\pmb v}\rangle$$
can be considered as if it were only 2 variables, 
where 
$\langle\tilde{\pmb a},\tilde{\pmb u}\rangle{\ctt}
\langle\tilde{\pmb a},\tilde{\pmb v}\rangle=
\frac{1}{2}\big(\langle\tilde{\pmb a},\tilde{\pmb u}\rangle
{*}\langle\tilde{\pmb a},\tilde{\pmb v}\rangle
 {+}\langle\tilde{\pmb a},\tilde{\pmb v}\rangle
{*}\langle\tilde{\pmb a},\tilde{\pmb u}\rangle\big)$
and 
$\langle\tilde{\pmb a},\tilde{\pmb u}\rangle_*^2$ etc means 
the product by $*$.
Let $D=\gamma^2-\alpha\beta$ be its discriminant.

Hence, we have 
\begin{equation}\label{connect}
{:}e_*^{\frac{t}{\hbar}\langle\tilde{\pmb a},\tilde{\pmb u}\rangle{\ctt}
\langle\tilde{\pmb a},\tilde{\pmb v}\rangle}{:}_{_{K_0}}
=e^{\frac{it}{2}}e^{\frac{i}{\hbar}(1-e^{it})
\langle\tilde{\pmb a},\tilde{\pmb u})\langle\tilde{\pmb a},\tilde{\pmb v}\rangle}.
\end{equation}
We define 
\begin{equation}\label{genpolar}
{\e}_{00}(\tilde{\pmb a})=
e_*^{\frac{\pi i}{i\h}\langle\tilde{\pmb a},\tilde{\pmb u}\rangle{\ctt}
\langle\tilde{\pmb a},\tilde{\pmb v}\rangle},\quad 
{:}{\e}_{00}(\tilde{\pmb a}){:}_{_{K_0}}=
ie^{-\frac{2}{i\h}
\langle\tilde{\pmb a},\tilde{\pmb u})\langle\tilde{\pmb a},\tilde{\pmb v}\rangle}
\end{equation}

We call $\e_{00}(\tilde{\pmb a})$ also a {\bf partial polar element}.
Since $[\langle\tilde{\pmb a},\tilde{\pmb u}\rangle, 
  \langle\tilde{\pmb b},\tilde{\pmb v}\rangle]=
-\hbar i\langle\tilde{\pmb a},\tilde{\pmb b}\rangle$, 
the product formula \eqref{eq:noteigi} gives  
$$ 
{:}\e_{00}(\tilde{\pmb a}){*}\e_{00}(\tilde{\pmb b}){:}_{_{K_0}}=
-e^{-\frac{2}{i\h}\big(
\langle\tilde{\pmb a},{\tilde{\pmb u}\rangle}
\langle\tilde{\pmb a},{\tilde{\pmb v}\rangle}
{+}\langle\tilde{\pmb b},{\tilde{\pmb u}\rangle}
  \langle\tilde{\pmb b},{\tilde{\pmb v}\rangle}
-2\langle\tilde{\pmb a},\tilde{\pmb b}\rangle
\langle\tilde{\pmb a},{\tilde{\pmb u}\rangle}
\langle\tilde{\pmb b},{\tilde{\pmb v}\rangle}\big)}
$$
In particular, if 
$\langle\tilde{\pmb a},\tilde{\pmb b}\rangle=0$, then  
$$
{\e}_{00}(\tilde{\pmb a}){*}\langle\tilde{\pmb b},\tilde{\pmb u}\rangle
=
\langle\tilde{\pmb b},\tilde{\pmb u}\rangle{*}{\e}_{00}(\tilde{\pmb a}),
\quad 
{\e}_{00}(\tilde{\pmb a}){*}\langle\tilde{\pmb b},\tilde{\pmb v}\rangle
=
\langle\tilde{\pmb b},\tilde{\pmb v}\rangle{*}{\e}_{00}(\tilde{\pmb a})
$$
and 
${:}\e_{00}(\tilde{\pmb a}){*}\e_{00}(\tilde{\pmb b}){:}_{_{K_0}}=
{:}\e_{00}(\tilde{\pmb b}){*}\e_{00}(\tilde{\pmb a}){:}_{_{K_0}}$.
By this observation, we have the following:
\begin{prop}\label{group}
  If 
$\langle\tilde{\pmb a},\tilde{\pmb b}\rangle=0$, then  
$$
{\e}_{00}(\tilde{\pmb a}){*}\langle\tilde{\pmb b},\tilde{\pmb u}\rangle
=
\langle\tilde{\pmb b},\tilde{\pmb u}\rangle{*}{\e}_{00}(\tilde{\pmb a}),
\quad 
{\e}_{00}(\tilde{\pmb a}){*}\langle\tilde{\pmb b},\tilde{\pmb v}\rangle
=
\langle\tilde{\pmb b},\tilde{\pmb v}\rangle{*}{\e}_{00}(\tilde{\pmb a})
$$
and 
$$
{:}\e_{00}(\tilde{\pmb a}){*}\e_{00}(\tilde{\pmb b}){:}_{_{K_0}}
={:}\e_{00}(\tilde{\pmb b}){*}\e_{00}(\tilde{\pmb a}){:}_{_{K_0}}
$$
\end{prop}

\bigskip
Let ${\mathfrak P}^{(\ell)}_{_{K_0}}$, $\ell=1,2,4$, be the set
consisting of all elements written by  
$$
{\e}_{00}(\tilde{\pmb a}_1){*}{\e}_{00}(\tilde{\pmb a}_2)
{*}\cdots{*}{\e}_{00}(\tilde{\pmb a}_{\ell k});
\quad k\in {\mathbb N},\,\,\tilde{\pmb a}_j\in {S}_{\mathbb C}^{m-1}.
$$
Since ${\e}_{00}(\tilde{\pmb a}_j)_*^{4}=1$, we have a series of subgroups 
${\mathfrak P}^{(1)}_{_{K_0}}\supset{\mathfrak P}^{(2)}_{_{K_0}}\supset
{\mathfrak P}^{(4)}_{_{K_0}}$ of ${\mathfrak G}$.  

\begin{lem}
${\mathfrak P}^{(4)}_{_{K_0}}$ is 
the group generated by 
$\{{\e}_{00}(\tilde{\pmb a}){*}{\e}^{-1}_{00}(\tilde{\pmb b}); 
\tilde{\pmb a},\tilde{\pmb b}\in S_{\mathbb C}^{m{-}1}\}$, 
which is a connected subgroup of ${\mathfrak G}$.   
However, $-1$ is not contained in  
${\mathfrak P}^{(4)}_{_{K_0}}$. 
Moreover 
${\mathfrak  P}^{(2)}_{_{K_0}}=
{\mathfrak P}^{(4)}_{_{K_0}}\bigcup 
({-}{\mathfrak P}^{(4)}_{_{K_0}})$, and 
$$
{\mathfrak  P}^{(1)}_{_{K_0}}=
\bigcup_{k=0}^{3}(i^k{\mathfrak P}^{(4)}_{_{K_0}})
$$  
\end{lem}

\noindent
{\bf Proof}\,\, It is easy to see that 
$\bigcup_{\ell}{\mathfrak  P}^{(4\ell)}_{_{K_0}}$ is a group.
Write its element as 
${\e}_{00}(\tilde{\pmb a}_1){*}
{\e}_{00}(\tilde{\pmb a}_2){*}\cdots{*}{\e}_{00}(\tilde{\pmb a}_{4\ell})$. 
Since ${S}_{\mathbb C}^{m-1}$ is arcwise connected, this is arcwise connected 
to $1^{\ell}=1$. 
For the second assertion, we have only to note that
 the product formula \eqref{eq:noteigi} allows that the phase part
and the amplitude part are computed independently.
\hfill $\Box$.

\subsubsection{Bumping identity}

For a while, we use notations $(u,v)$ for $(\tilde{u},\tilde{v})$ for simplicity. 
Using the uniqueness of the real analytic solution, 
we have the following useful 
\begin{lem}{\rm(bumping identity)}\label{bump}
$v*e_*^{it u*v}= 
e_*^{it v*u}*v$,\,\, $e_*^{it u*v}*u= u*e_*^{it v*u}$ 
holds.
\end{lem}

\noindent
{\bf Proof}\,\,This is given by 
the bumping identity $u*(v*u)^m=(u*v)^m*u$,  
if the polynomial approximation theorem holds.
Here, this is proved by the uniqueness of the 
real analytic solution, hence the proof can be applied 
for other expressions. The continuity of $v*$ and 
the associativity give 
$$
\frac{d}{dt}v*e_*^{it u*v}
=v*((iu*v)*e_*^{it u*v})=iv*u*(v*e_*^{it u*v}).
$$
On the other hand, the continuity of  
$*v$ and the associativity give 
$$
\frac{d}{dt}e_*^{it v*u}*v =
((iv*u)*e_*^{it u*v})*v=iv*u*(e_*^{it u*v}*v).
$$
Both satisfy the differential equation 
$\frac{d}{dt}f_t=iv*u*f_t$ with the initial condition $v$.
The second one is shown by the same proof. 
 \hfill$\Box$  

Note here that we used only the real analyticity of 
$e_*^{it u*v}$ in the normal ordered expression. Note that the bumping
identiy gives in particular 
$$
v*e_*^{\frac{\pi i}{i\h}u{\ctt}v}
{=}e_*^{\frac{\pi i}{i\h}(u{\ctt}v{+}i\h)}*v
{=}{-}e_*^{\frac{\pi i}{i\h}u{\ctt}v}{*}v,\quad  
u*e_*^{\frac{\pi i}{i\h}u{\ctt}v}
{=}{-}e_*^{\frac{\pi i}{i\h}u{\ctt}v}{*}u.
$$ 

The next Proposition is stated under $K_0$-expression by using the
original notations, but we see 
the bumping identity holds in every ordered expression defined 
in the next section \S\,\ref{PBW-theorem}.   
\begin{prop}\label{bump33}
${:}\langle\tilde{\pmb a},\tilde{\pmb u}\rangle{*}\e_{00}(\tilde{\pmb  a}){:}_{_{K_0}}
=
-{:}\e_{00}(\tilde{\pmb  a}){*}\langle\tilde{\pmb a},\tilde{\pmb u}\rangle{:}_{_{K_0}}$
and \,\,
${:}\langle\tilde{\pmb a},\tilde{\pmb v}\rangle{*}\e_{00}(\tilde{\pmb  a}){:}_{_{K_0}}
=
-{:}\e_{00}(\tilde{\pmb  a}){*}\langle\tilde{\pmb a},\tilde{\pmb v}\rangle{:}_{_{K_0}}$.
\end{prop}

\subsection{Double covering group of $SO(m,{\mathbb C})$ 
in ${\mathbb C}_*{\times}G\!L(m,{\mathbb C})$ } 


Our main concern in this section is the mutual relations between
$*$-exponential functions of degenerate quadratic forms of small rank.   
  
By Propositions\,\,\ref{group} and  \ref{bump33}, we have easily that  
$$
e_*^{\frac{\pi}{\hbar}\tilde u_1{\ctt}\tilde v_1}{*}(\sum^m_{i=1}b_i\tilde u_i)
{*}e_*^{-\frac{\pi}{\hbar}\tilde u_1{\ctt}\tilde v_1}
=-b_1\tilde u_1+\sum^m_{i=2}b_i\tilde u_i. 
$$


\medskip
In general, for every 
$\langle\tilde{\pmb a},\tilde{\pmb u}\rangle, \tilde{\pmb a}\in 
S_{\mathbb C}^{m-1}$ and $\tilde{\pmb b}\in{\mathbb C}^{m}$, 
we set 
$\tilde{\pmb b}=
\langle\tilde{\pmb a},\tilde{\pmb b}\rangle\tilde{\pmb a}{+}\tilde{\pmb c},
\,\, \langle\tilde{\pmb a},\tilde{\pmb c}\rangle=0$.   
By computing the same way as above,  we have the reflection with respect to 
$\tilde{\pmb a}$:  
\begin{equation}
  \label{eq:orikaesi}
\begin{aligned}
e_*^{\frac{\pi}{\hbar}\langle\tilde{\pmb a},\tilde{\pmb u}\rangle{\ctt}
\langle\tilde{\pmb a},\tilde{\pmb v})}*\langle\tilde{\pmb b},\tilde{\pmb u}\rangle
*e_*^{-\frac{\pi}{\hbar}\langle\tilde{\pmb a},\tilde{\pmb u}){\ctt}
\langle\tilde{\pmb a},\tilde{\pmb v}\rangle}
=&\langle\tilde{\pmb b}{-}2\langle\tilde{\pmb a},\tilde{\pmb b}\rangle\tilde{\pmb a},\,\,\tilde{\pmb u}\rangle\\
e_*^{\frac{\pi}{\hbar}\langle\tilde{\pmb a},\tilde{\pmb u}\rangle{\ctt}
\langle\tilde{\pmb a},\tilde{\pmb v})}*\langle\tilde{\pmb b},\tilde{\pmb v}\rangle
*e_*^{-\frac{\pi}{\hbar}\langle\tilde{\pmb a},\tilde{\pmb u}){\ctt}
\langle\tilde{\pmb a},\tilde{\pmb v}\rangle}
=&\langle\tilde{\pmb b}{-}2\langle\tilde{\pmb a},\tilde{\pmb b}\rangle\tilde{\pmb a},\,\, \tilde{\pmb v}\rangle\\
\end{aligned}
\end{equation}
For simplicity we use notations in the group theory 
$$
{\rm{Ad}}(\e_{00}(\tilde{\pmb a}))X=
\e_{00}(\tilde{\pmb a}){*}X{*}\e_{00}^{-1}(\tilde{\pmb a}),\quad 
{\rm{Ad}}(\e_{00}(\tilde{\pmb a}){*}\e_{00}(\tilde{\pmb b}))X=
\e_{00}(\tilde{\pmb a}){*}\e_{00}(\tilde{\pmb b}){*}X{*}
\e_{00}^{-1}(\tilde{\pmb b}){*}\e_{00}^{-1}(\tilde{\pmb a}).
$$
Note that ${\rm{Ad}}(\e_{00}(\tilde{\pmb a}))={\rm{Ad}}(\e_{00}^{-1}(\tilde{\pmb a}))$.

Now, Proposition\,\ref{group} gives the following   
\begin{prop}\label{Clifford}
${\rm{Ad}}(\e_{00}(\tilde{\pmb a}){*}\e_{00}^{-1}(\tilde{\pmb b}))$ 
generate $SO(m,\mathbb C)$.  
Since ${\mathfrak P}^{(4)}_{_{K_0}}$ does not contain $-1$, 
$$
\{\e_{00}(\tilde{\pmb a}){*}\e_{00}^{-1}(\tilde{\pmb b}); 
\tilde{\pmb a}, \tilde{\pmb b}\in {S}^{m{-}1}_{\mathbb C}\}
$$  
generates a group isomorphic to $SO(m,\mathbb C)$. 
If $\tilde{\pmb a}, \tilde{\pmb b}$ are
restricted in real vectors, then 
${\rm{Ad}}(\e_{00}(\tilde{\pmb a}){*}\e_{00}^{-1}(\tilde{\pmb b}))$ generate $SO(m)$, hence 
$\{\e_{00}(\tilde{\pmb a}){*}\e_{00}^{-1}(\tilde{\pmb b})\}$ 
generate a group isomorphic to $SO(m)$.  
\end{prop}

\medskip
Since  
$$
{\rm{Ad}}(\e_{00}(\tilde{\pmb a}){*}\e_{00}(\tilde{\pmb b}))=
{\rm{Ad}}(\e_{00}(\tilde{\pmb a}){*}\e_{00}^{-1}(\tilde{\pmb b}))
$$
we conclude the following:
\begin{thm}\label{polargroup}
In the normal ordered expression, 
$\{e_*^{\frac{\pi}{\h}(\tilde{\pmb a},\tilde{\pmb u}){\ctt}
(\tilde{\pmb a},\tilde{\pmb v})}{*}e_*^{\frac{\pi}{\h}(\tilde{\pmb b},\tilde{\pmb u}){\ctt}
(\tilde{\pmb b},\tilde{\pmb v})}; 
\tilde{\pmb a}, \tilde{\pmb b}\in S_{\mathbb C}^{m-1} \}$ 
forms a group  $\widetilde{SO}(m,\mathbb C)$ which is a 
subgroup of ${\mathfrak G}$, and 
${\rm{Ad}}(\widetilde{SO}(m,\mathbb C))=SO(m,\mathbb C)$. 
$$
{\rm{Ad}}: \widetilde{SO}(m,\mathbb C)\to SO(m,{\mathbb C})
$$
is a  $2$-to-$1$ surjective homomorphism.

Moreover, the mapping $\pi$ given by forgetting the amplitude part 
gives also a  $2$-to-$1$ homomorphism onto a subgroup of 
${\mathcal O}'_{_{K_0}}$, which will be denoted by ${SO}_{*}(m,{\mathbb C})$. 
\end{thm}

\setlength{\unitlength}{.5mm}
\begin{picture}(60,90)
\put(80,60){${SO}_{*}(m,{\mathbb C})$}
\put(35,63){\vector(1,0){35}}
\put(40,65){\footnotesize{forget amp.}}
\put(0,60){$\widetilde{SO}(m,\mathbb C)$}
\put(85,50){\vector(0,-1){30}}
\put(80,10){${SO}(m,{\mathbb C})$}
\put(30,55){\vector(1,-1){30}}
\put(30,35){Ad}
\put(85,35){{Ad}'}
\end{picture}  
\hfill
{\parbox[b]{.5\linewidth}
{ It is not hard to 
consider the adjoint action of the group 
${SO}_{*}(m,{\mathbb C})$. This is denoted by ${\rm{Ad}'}$. 
This is an isomorphism. 
It is remarkable that Theorem\,\ref{polargroup} is obtained without using 
Clifford algebra. However, if the obtained group 
$\widetilde{SO}(m)=Spin(m)$ is constructed via Clifford algebra, 
we must have 
$(\e_{00}(\tilde{\pmb a}){*}\e_{00}(\tilde{\pmb b}))_*^2=-1$, but 
such an anti-commutative property is not given by 
the normal ordered expression.} 

Hence, we have in normal ordered expression that 
$$
\widetilde{SO}(m,\mathbb C)\cong{SO}(m,\mathbb C){\times}{\mathbb Z}_2,\quad
\widetilde{SO}(m)\cong {SO}(m){\times}{\mathbb Z}_2.
$$
In the next subsection, we prove some strange nature of partial polar elements
in the normal ordered expression.

\subsection{Partial polar elements are double-valued}\label{anomalousexp}

For every $\tilde{\pmb a}\in {S}_{\mathbb C}^{m-1}$, we consider   
$$
\e_{00}(\tilde{\pmb a})=
e_*^{\frac{\pi}{\hbar}(\tilde{\pmb a},\tilde{\pmb u})
{\ctt}(\tilde{\pmb a},\tilde{\pmb v})},
\quad\,\,
\e_{00}^{-1}(\tilde{\pmb a})=
e_*^{-\frac{\pi}{\hbar}(\tilde{\pmb a},\tilde{\pmb u})
{\ctt}(\tilde{\pmb a},\tilde{\pmb v})}. 
$$

If 
$\langle\tilde{\pmb a},\tilde{\pmb a}\rangle=1$, then 
$\langle\tilde{\pmb a},\tilde{\pmb u}\rangle$ and 
$\langle\tilde{\pmb a},\tilde{\pmb v}\rangle$ form a canonical
conjugate pair. In this subsection, we set 
$u=\langle\tilde{\pmb a},\tilde{\pmb u}\rangle$ and 
$v=\langle\tilde{\pmb a},\tilde{\pmb v}\rangle$ for simplicity. 

In the normal ordered expression  $K_0$, we set 
$$
{:}e_*^{{\frac{t}{\hbar}}(au^2+bv^2+2cu*v)}{:}_{_{K_0}}=  
  \psi(t)e^{X(t)u^2+Y(t)v^2+2Z(t)uv}.
$$   
then we have a system of ordinary differential equations with initial
conditions 
$X(0)=Y(0)=Z(0)=0$ and $\psi(0)=1$
  \begin{equation}
    \label{eq:diffeq8}
 \left\{\begin{aligned}
    X'(t)=&\frac{1}{\hbar}a+4icX(t)-4\hbar bX(t)^2\\
    Y'(t)=&\frac{1}{\hbar}b+ 4ibZ(t)-4\hbar bZ(t)^2\\
    Z'(t)=&\frac{1}{\hbar}c+2icZ(t)+2ibX(t)-4\hbar bX(t)Z(t)\\
    \psi'(t)  =&-2\hbar bX(t)\psi(t)
  \end{aligned}\right.
  \end{equation}

Solving the evolution equation \eqref{eq:diffeq8}, we have the
solution \eqref{eq:4.14} for $D{=}c^2\!-\!ab=1$. 
 
\begin{equation}\label{eq:4.14}
\left\{\begin{aligned}
X(t)&=\frac{a}{2}\frac{\sin (2t)}
    {\cos(2t)-ic\sin (2t)} \\
Y(t)&=\frac{b}{2}\frac{\sin (2t)}
    {\cos(2t)-ic\sin(2t)},\\
Z(t)&=\frac{i}{2}\big(1-\frac{1}{\cos (2t)-ic\sin (2t)}\big) \\
\psi(t)&=\frac{e^{-itc}}{\sqrt{\cos(2t)-ic\sin(2t)}}
\end{aligned}
\right.
\end{equation} 
Readers have only to check this is a real analytic solution, and need
not to care about how this form is obtained. 

Although 
$e_*^{\frac{\pi i}{2i\h}(au^2+bv^2+c(u\!*\!v+v\!*\!u))}$ 
diverges in the Weyl ordered expression, recall that  
its normal ordered expression has been given in the previous section.

\medskip
\noindent
{\bf Strange behaviour of solutions}\,\, 

Here, we mention an anomalous behaviour of the solution in  
the normal ordered expression. 
We think these have never emphasized in physics literatures, 
but these are crucial from our view point. 

\noindent
({\bf a})\,\, It is remarkable that if $c{\not=}0$, e.g $c=\pm 1$,  
then $\sqrt{\cos(2t)-ic\sin(2t)}$ must change sign  
on $[0,\pi]$, since the curve turning around the 
origin. Thus, one has to set 
$\sqrt{\cos(2\pi)-ic\sin(2\pi)}=-1$, whenever 
$\sqrt{\cos(0)-ic\sin(0)}=1$ is needed. Thus, we have 
\begin{equation}
{:}e_*^{\frac{\pi i}{i\h}(au^2+bv^2+2cu{*}v)}{:}_{_{K_0}}=
-e^{-\pi ic}
\end{equation} 
depending only on $c$ whenever $c^2{-}ab=1$.  

\medskip
\noindent
({\bf b})\,\,
Branched singular points are distributed $\pi$-periodically along two lines  parallel to 
the real line lying in upper or lower  
half-plane depending on $c$.  
On the other hand, if $c=0$, 
$\sqrt{\cos(2t)}$ has two singular 
points at $t{=}\frac{\pi}{4}, \frac{3\pi}{4}$.

\medskip
\noindent
({\bf c})\,\,Along the pure-imaginary direction, if $|{\rm{Re}}\,c|<1$, then 
${:}e_*^{\frac{t}{i\h}(au^2+bv^2+2cu{*}v)}{:}_{_{K_0}}$ is rapidly
decreasing in both sides. Hence, the integral 
$\int_{\mathbb R}{:}e_*^{\frac{t}{i\h}(au^2+bv^2+2cu{*}v)}{:}_{_{K_0}}$ 
converges. 
If ${\rm{Re}}\,c{=}1$,
then we have nonvanishing limits 
$$
\lim_{t\to\infty}
{:}e_*^{\frac{t}{i\h}(au^2+bv^2+2cu{*}v)}{:}_{_{K_0}}, \quad 
\lim_{t\to-\infty} 
{:}e_*^{\frac{t}{i\h}(au^2+bv^2+2cv{*}u)}{:}_{_{K_0}}.
$$
These limits play the role of ground states (vacuums) in  matrix
representations (cf.\eqref{vacuua}).    

\bigskip 
The most strange phenomenon will be seen below: 

\noindent
({\bf d})\,\,
Set $u{\ctt}v{=}\frac{1}{2}(u{*}v{+}v{*}u)$. 
By noting that $2u{\ctt}v{=}2u{*}v{+}i\h$, 
\eqref{eq:4.14} shows that the term $e^{{-}itc}$
disappear in the normal ordered expression of 
$e_*^{\frac{t}{\h}(au^2+bv^2+2cu{\ctt}v)}$, 
and hence at $t=\pi$, we have 
$$
{:}e_*^{\frac{\pi i}{i\h}(au^2+bv^2+2cu{\ctt}v)}{:}_{_{K_0}}
=-1
$$
independent of $a,b,c$ whenever $c^2-ab=1$. Moreover, at 
$t=\frac{\pi}{2}$ we have  
$$
{:}e_*^{\frac{\pi}{2\h}(au^2+bv^2+2cu{\ctt}v)}{:}_{_{K_0}}{=}
\sqrt{-1}e^{-\frac{2}{i\h}uv} 
$$
independent of $a, b, c$ whenever $c^2{-}ab{=}1$.  
Since the manifold $\{c^2{-}ab{=}1\}$ is arcwise connected, 
the sign ambiguity of $\sqrt{-1}$ must be eliminated, and 
$$
\big\{{:}e_*^{\frac{\pi}{2\h}
(au^2+bv^2+2cu{\ctt}v)}{:}_{_{K_0}};
a, b, c {\in}{\mathbb C},\, \,c^2{-}ab{=}1\big\}
$$
must be viewed as a single element. In particular, since  
$(a,b,c)=(0,0,1)$ and $(0,0,-1)$ are arcwise connected 
in the set $c^2{-}ab{=}1$, we have  
\begin{equation}
  \label{eq:contra}
{:}e_*^{\frac{\pi}{2\h}2u{\ctt}v}{:}_{_{K_0}}=
\sqrt{-1}e^{-\frac{2}{i\h}uv}=
{:}e_*^{-\frac{\pi}{2\h}2u{\ctt}v}{:}_{_{K_0}}
\end{equation} 

\bigskip

Recall that the 
$*$-exponential function 
${:}e_*^{\frac{ti}{i\h}2u{\ctt}v}{:}_{_{K_0}}$ 
is holomorphic for $t\in {\mathbb C}$ and  
by the definition of the exponential function, we must set 
${:}e_*^{\frac{\pi}{\h}2u{\ctt}v}{:}_{_{K_0}}={-}1$ and 
${:}e_*^{\frac{\pi}{2\h}2u{\ctt}v}{:}_{_{K_0}}
=ie^{-\frac{2}{i\h}uv}$ by 
fixing $1$ at $t=0$. The exponential law gives 
\begin{equation}\label{anomary}
{:}e_*^{\frac{\pi}{2\h}2u{\ctt}v}
{*}e_*^{\frac{\pi}{2\h}2u{\ctt}v}{:}_{_{K_0}}
={:}e_*^{\frac{\pi}{\h}2u{\ctt}v}{:}_{_{K_0}}=-1. 
\end{equation}
Hence, we have the anomalous identity 
$$
{:}e_*^{\frac{\pi}{2\h}2u{\ctt}v}{:}_{_{K_0}}=
-{:}e_*^{-\frac{\pi}{2\h}2u{\ctt}v}{:}_{_{K_0}},
$$
which appears to contradict to \eqref{eq:contra}. Note that 
the exponential law is established by the uniqueness of the real 
analytic solution of the evolution equation.

Indeed, this is not a contradiction, but 
${:}e_*^{\frac{\pi}{2\h}2u{\ctt}v}{:}_{_{K_0}}$ is  
{\it a double-valued single element} caused since 
the ambiguity of $\sqrt{-1}$ cannot be removed.   
We called it the {\bf polar element} and denote 
it by ${\e}_{00}$. ${\e}_{00}$ is an element 
something like $\sqrt{-1}$, or an operator 
whose eigenvalues are $\pm i$. 

\bigskip
In fact, this pathological phenomenon is caused by 
singular points of 
${:}e_*^{\frac{t}{\hbar}(au^2+bv^2+2cu{\ctt}v)}{:}_{_{K_0}}$
lying in the domain 
$\{(t,a,b,c); 0{<}t{<}\frac{\pi}{2}, c^2{-}ab{=}1\}$
(cf. \cite{ommy}, \cite{OMMY6}). This is just like 
$\frac{1}{\sqrt{\cos(2t)}}$ in ({\bf b}) above.

\bigskip
To explain this, we first note that
$[\frac{1}{i\h}(u^2{+}v^2)_*,u]{=}2v,\,\, 
[\frac{1}{i\h}(u^2{+}v^2)_*,v]{=}{-}2u$. 
By this we have for each $\theta$, one parameter groups  
Ad$(e_*^{\frac{i\theta}{2\hbar}(u^2+v^2)})e_*^{2tu{\ctt}v}$
with respect to $t$:
$$
{\rm{Ad}}(e_*^{\frac{i\theta}{2\h}(u^2+v^2)})e_*^{2tu{\ctt}v}
=e_*^{t(\sin 2\theta\,\,(u^2-v^2)+\cos 2\theta \,\,2u{\ctt}v)}, \quad 
{\rm{Ad}}(e_*^{\frac{\pi i}{4\hbar}(u^2+v^2)})e_*^{2tu{\ctt}v}=e_{*}^{-2tu{\ctt}v}.
$$ 
In particular, setting $t=\frac{\pi}{2\h}$ we have   
$$
{\rm{Ad}}(e_*^{\frac{\pi i}{4\hbar}(u^2+v^2)})e_*^{\frac{\pi}{\h}u{\ctt}v}=e_{*}^{-\frac{\pi}{\h}u{\ctt}v}
$$

On the other hand, if we fix $t{=}\frac{\pi}{2}$ first, 
then the element 
${:}{\rm{Ad}}
(e_*^{\frac{\theta i}{4\hbar}(u^2+v^2)})\e_{00}{:}_{_{K_0}}$ is
independent of $\theta$ by the formula mentioned in ({\bf d}), since the discriminant of the quadratic form of 
the r.h.s. is identically $1$. The normal ordered expression of  
the r.h.s. is identically ${:}e_*^{\frac{\pi}{\h}u{\ctt}v}{:}_{_{K_0}}$ for 
$t=\frac{\pi}{2\h}$, that is,  
$$
{\rm{Ad}}(e_*^{\frac{i\theta}{2\h}(u^2+v^2)})e_*^{\frac{\pi}{\h}u{\ctt}v}=e_*^{\frac{\pi}{\h}u{\ctt}v}.
$$
In fact, the r.h.s. $e_*^{\frac{\pi}{\h}u{\ctt}v}$ is not on the exponential function but 
on another ``exponential function'' in the opposite sheet  which is 
${-}1$ at $t=0$. 
Exchanging sheet is caused by the branched singular point at 
$2\theta=\frac{\pi}{2}$.

\medskip
\noindent
{\bf Remark}\,\, Although examples in this section  
are stated in the normal ordered expression for various 
quadratic forms, \eqref{eq:KK2} in the next section shows that same phenomena 
must appear for a single quadratic form $\frac{1}{i\h}u{\ctt}v$ 
under various expression parameters. 

\section{General product formulas and intertwiners}\label{PBW-theorem} 

To understand strange phenomena mentioned in the previous section, we
are requested to make a wider view about expression for elements of    
transcendentally extended Weyl algebra. 

For that, we start in a general setting as follows:
Let ${\mathfrak S}(n)$ and ${\mathfrak A}(n)$ be 
the spaces of complex symmetric matrices and skew-symmetric 
matrices respectively, and 
${\mathfrak M}(n){=}
{\mathfrak S}(n)\oplus{\mathfrak A}(n)$.
For an arbitrary fixed $n{\times}n$-complex matrix 
$\Lambda{\in}{\mathfrak M}(n)$, we 
define a product ${*}_{_{\Lambda}}$ on the space of polynomials   
${\mathbb C}[\pmb u]$ by the formula 
\begin{equation}
 \label{eq:KK}
 f*_{_{\Lambda}}g=fe^{\frac{i\h}{2}
(\sum\overleftarrow{\partial_{u_i}}
{\Lambda}_{ij}\overrightarrow{\partial_{u_j}})}g
=\sum_{k}\frac{(i\h)^k}{k!2^k}
{\Lambda}_{i_1j_1}\!{\cdots}{\Lambda}_{i_kj_k}
\partial_{u_{i_1}}\!{\cdots}\partial_{u_{i_k}}f\,\,
\partial_{u_{j_1}}\!{\cdots}\partial_{u_{j_k}}g.   
\end{equation}
It is known and not hard to prove that 
$({\mathbb C}[\pmb u],*_{_{\Lambda}})$ is an associative algebra. 
Clearly, if $\Lambda$ is symmetric, then the obtained algebra 
is commutative and it is isomorphic  
to the standard polynomial algebra with $\h$. 

\medskip
For every $\Lambda$, $\partial_{u_{i}}$ acts as a 
derivation of the algebra 
$({\mathbb C}[\pmb u],*_{_{\Lambda}})$. 
Noting this, for any other constant symmetric matrix $K$,
define a new product $*_{_{\Lambda,K}}$ by the formula  
$$  
\begin{aligned}
f*_{_{\Lambda{,}K}}g=&f
e^{\frac{i\h}{2}
(\sum\overleftarrow{\partial_{u_i}}
{K}_{ij}{*_{_{\Lambda}}}\overrightarrow{\partial_{u_j}})}g\\
=&\sum_{k}\frac{(i\h)^k}{k!2^k}
{K}_{i_1j_1}\cdots{K}_{i_kj_k}
(\partial_{u_{i_1}}\cdots\partial_{u_{i_k}}f){*_{_{\Lambda}}}
(\partial_{u_{j_1}}\cdots\partial_{u_{j_k}}g). 
\end{aligned}
$$
This is  also an associative algebra 
$({\mathbb C}[\pmb u],*_{_{\Lambda,K}})$. 
Since $\Lambda$, $K$ are constant matrices 
and the non-commutativity of matrix algebra 
is not used in calculation of product formulas, 
the new product formula can be rewritten as 
$$
f*_{_{\Lambda,K}}g=
\sum_{k}\frac{(i\h)^k}{k!2^k}
{(\Lambda{+}K)}_{i_1j_1}
\cdots{(\Lambda{+}K)}_{i_kj_k}
\partial_{u_{i_1}}\cdots\partial_{u_{i_k}}f
\partial_{u_{j_1}}\cdots\partial_{u_{j_k}}g
$$ 
by noting that exchanging indeces of 
$\partial_{u_{i_1}\cdots u_{i_k}}$ is permitted. 
That is, ${*}_{_{\Lambda,K}}={*}_{_{\Lambda+K}}$.

This formula may be written as
\begin{equation}
 \label{eq:prprod}
fe^{\frac{i\h}{2}(\sum\overleftarrow{\partial_{u_i}}
{(\Lambda{+}K)}_{ij}\overrightarrow{\partial_{u_j}})}g
=
fe^{\frac{i\h}{2}(\sum\overleftarrow{\partial_{u_i}}
{K}_{ij}
{e^{\frac{i\h}{2}(\sum\overleftarrow{\partial_{u_k}}
{\Lambda}_{kl}\overrightarrow{\partial_{u_k}})}}
  \overrightarrow{\partial_{u_j}})}g.
\end{equation}

Using a symmetric matrix $K$, we compute 
$\frac{1}{k!}
(\frac{i\h}{4}\sum{K}_{ij}
\partial_{u_i}\partial_{u_j})^k(f{*}_{_K}g)$ 
by noting that this is written as follows:
$$
\sum_{p{+}q{+}r=k}
\frac{(i\h)^r}{r!2^r}
{K}_{i_1j_1}\cdots{K}_{i_rj_r}
\partial_{u_{i_1}}\cdots\partial_{u_{i_r}}
\frac{1}{p!}(\frac{i\h}{4}
\sum{K}_{ij}\partial_{u_i}\partial_{u_j})^pf
\,\,\,
\partial_{u_{j_1}}\cdots\partial_{u_{j_r}}
\frac{1}{q!}
(\frac{i\h}{4}\sum{K}_{ij}\partial_{u_i}\partial_{u_j})^qg.
$$
Using this formula, we have the following formula:
\begin{equation}
  \label{eq:Hochsch}
  \begin{aligned}
e^{\frac{i\h}{4}\sum{K}_{ij}\partial_{u_i}\partial_{u_j}}
\Big(&\big(e^{-\frac{i\h}{4}\sum{K}_{ij}
\partial_{u_i}\partial_{u_j}}f\big)
{*_{_\Lambda}}
\big(e^{-\frac{i\h}{4}\sum{K}_{ij}\partial_{u_i}\partial_{u_j}}g\big)\Big)\\
=&fe^{\frac{i\h}{2}(\sum\overleftarrow{\partial_{u_i}}
{*_{\Lambda}}{K}_{ij}{*_{_\Lambda}}
\overrightarrow{\partial_{u_j}})}g= f{*}_{_{\Lambda{+}K}}g.
 \end{aligned}
\end{equation}

The next one is proved directly by formula \eqref{eq:Hochsch}:
\begin{cor}
Let 
$I_0^{^K}(f)=
e^{\frac{i\h}{4}\sum{K}_{ij}\partial_{u_i}\partial_{u_j}}$ 
and $I_{_K}^0(f)=
e^{-\frac{i\h}{4}\sum{K}_{ij}\partial_{u_i}\partial_{u_j}}$. 
Then $I_0^{^K}$ is an isomorphism of 
$({\mathbb C}[\pmb u];{*}_{_{\Lambda}})$ onto 
$({\mathbb C}[\pmb u];{*}_{_{\Lambda+K}})$.
\end{cor}
Set $\Lambda=K{+}J$ where $K$, $J$ are the symmetric part
and the skew part of $\Lambda$, respectively. 
Since the commutator 
$[u_i,u_j]={i\h}J_{ij}$ is given by the skew part of $\Lambda$, 
the algebraic structure of 
$({\mathbb C}[\pmb u], *_{_\Lambda})$ depends only on $J$, whose 
isomorphism class may be denoted by $({\mathbb C}[\pmb u], *_{_J})$
or simply by $({\mathbb C}[\pmb u], *)$ by noticing this class 
consists of a {\it single} algebra.   

\medskip
It is clear that the product $f{*_{_{\Lambda}}}g$ 
is defined if one of $f, g$ is 
a polynomial and another is a smooth function.

Let $H{\!o}l({\mathbb C}^n)$ be the space of all 
holomorphic functions on the complex $n$-plane ${\mathbb C}^n$ with 
the uniform convergence topology on each compact domain. 
The following two propositions are useful:
\begin{prop}\label{frechet} 
$H{\!o}l({\mathbb C}^n)$ with the topology above 
is a Fr{\'e}chet space defined by a countable family of seminorms. 
\end{prop}

\begin{prop}\label{extholom} 
For every polynomial $p(\pmb u)\in{\mathbb C}[\pmb u]$, 
the left-multiplication 
$f\to p(\pmb u){*_{_{\Lambda}}}f$ and the right-multiplication 
$f\to f{*_{_{\Lambda}}}p(\pmb u)$ are both continuous 
linear mappings of $H\!ol({\mathbb C}^n)$ into itself. 

If two of $f, g, h$ are polynomials, then associativity 
$(f{*_{_{\Lambda}}}g{*_{_{\Lambda}}})h=
f{*_{_{\Lambda}}}(g{*_{_{\Lambda}}}h)$
holds.
\end{prop}

\subsection{Expression parameters and intertwiners}
\label{Expinter}
As used in the previous section, we recall notations 
\begin{equation}\label{Weyl}
{\pmb u}=(u_1,u_2,\cdots,u_{2m})=
(\tilde{\pmb u},\tilde{\pmb v}),\quad  
\tilde{\pmb u}=(\tilde{u}_1,\cdots,\tilde{u}_m),\,\,
\tilde{\pmb v}=(\tilde{v}_1,\cdots,\tilde{v}_m).
\end{equation}
The skew part $J$ is fixed to be the standard skew-symmetric 
matrix 
$J=\left[
{\footnotesize 
{\begin{matrix} 
   0 & {-}I\\ 
   I & 0 
 \end{matrix}}} 
\right]$. The algebra is called 
the {\bf Weyl algebra} and the isomorphism class is 
denoted by $W_{\h}(2m)$. 

In the standard theory of algebraic system, 
we are only concerned with isomorphism 
classes. For the case of a universal enveloping algebra 
of a Lie algebra, Poincar{\'e}-Birkhoff-Witt theorem ensures
that this is realized on the space of ordinary 
polynomials by giving a new associative product. 
However, there is no standard way of 
unique expression of elements for algebras. 

Note that if the generator system is fixed, 
then the product formula \eqref{eq:KK} 
also gives the unique expression of elements of 
this algebra by the ordinary polynomials. 

\medskip
For instance, computing $u_i{*}u_j{da*}u_k$ using 
\eqref{eq:KK} gives the 
expression of $u_i{*}u_j{*}u_k$ as a polynomial. Thus, 
the product formula
\eqref{eq:KK} will be referred as a $K$-{\it ordered expression}, 
i.e. if generators are fixed, giving an ordered expression 
is nothing but giving a product formula on the space of 
polynomials which defines the Weyl algebra $W_{\h}(2m)$.

\medskip
By this formulation of orderings, the intertwiner between
$K$-ordered expressions and $K'$-ordered expressions 
is explicitly given as follows:
\begin{prop}
\label{intwn}
For every $K, K'\in{\mathfrak S}(n)$, the intertwiner
is defined by  
\begin{equation}
\label{intertwiner}
I_{_K}^{^{K'}}(f)=
\exp\Big(\frac{i\h}{4}\sum_{i,j}(K'_{ij}{-}K_{ij})
\partial_{u_i}\partial_{u_j}\Big)f \,\,
(=I_{0}^{^{K'}}(I_{0}^{^{K}})^{-1}(f)), 
\end{equation}
and by \eqref{eq:Hochsch}, it gives an isomorphism 
$I_{_K}^{^{K'}}:({\mathbb C}[{\pmb u}]; *_{_{K+J}})\rightarrow 
({\mathbb C}[{\pmb u}]; *_{_{K'+J}})$.
Namely, the following identity holds for any 
$f,g \in {\mathbb C}[{\pmb u}]:$ 
\begin{equation}\label{intertwiner2}
I_{_K}^{^{K'}}(f*_{_\Lambda}g)=
I_{_K}^{^{K'}}(f)*_{_{\Lambda'}}I_{_K}^{^{K'}}(g),
\end{equation}
where $\Lambda=K{+}J$, $\Lambda'=K'{+}J$.
\end{prop}
Intertwiners do not change the algebraic structure $*$, 
but do change the expression of elements by the ordinary 
commutative structure.  

\begin{center}
\fbox{If the skew part $J$ is fixed, 
we often use notation $*_{_K}$ instead of $*_{_\Lambda}$}
\end{center}

As in the case of one variable, infinitesimal intertwiner 
$$
dI_{_K}(K')=\frac{d}{dt}\Big|_{t=0}I_{_K}^{^{K{+}tK'}}=\frac{1}{4i\h}{K'}_{ij}\partial_{u_i}\partial_{u_j}
$$
is viewed as a flat connection on the trivial bundle 
$\coprod_{K\in{\mathfrak S}(n)}
H{\!o}l({\Bbb C}^{n})$. The equation of parallel translation 
along a curve $K(t)$ is given by 
\begin{equation}\label{parallel}
\frac{d}{dt}f_t=
dI_{_K}(\dot K(t))f_t, \quad 
\dot K(t)=\frac{d}{dt}K(t),
\end{equation}
but this may not have a solution for some initial function.

\subsubsection{Several remarks on product 
formulas and notations}\label{formula00}

In what follows, we use the notation $*_{_K}$ instead of 
$*_{_\Lambda}$, since the skew part $J$ is fixed as the 
standard skew-matrix.
We use notations 
$$
{\pmb u}=(u_1,u_2,\cdots,u_{2m})=
(\tilde{\pmb u},\tilde{\pmb v}),\quad  
\tilde{\pmb u}=(\tilde{u}_1,\cdots,\tilde{u}_m),\,\,
\tilde{\pmb v}=(\tilde{v}_1,\cdots,\tilde{v}_m).
$$

All results in \cite{OMMY3} hold for functions  
$f(\langle{\pmb a},{\pmb u}\rangle)$ by setting 
$\tau=i\h\langle{\pmb a}K,{\pmb a}\rangle$.

Note that according to the choice of $K=0, K_0, {-}K_0$, $I$,  
where 
$$
(0,\,\, K_0, {-}K_0, I)= 
\left(
 \left[
{\footnotesize
{\begin{matrix}
   0 & 0\\
   0 & 0
\end{matrix}}}
 \right],\,\,
\left[
{\footnotesize
{\begin{matrix}
   0 & I\\
   I & 0
 \end{matrix}}}
\right],\,\,
\left[
{\footnotesize
{\begin{matrix}
   0 &\!\!{-}I\\
   {-}I&\!\!0
 \end{matrix}}}\right],\,\,
\left[
{\footnotesize
{\begin{matrix}
   I&\!\!0\\
   0&\!\!I
 \end{matrix}}}\right]
\right),
$$
\begin{tabular}{l|l} 
Choice of $K$ & (name of ordering)\\ \hline
$K=0$         & Weyl ordered expression\\ \hline 
$K_0=\left[
{\footnotesize{
\begin{matrix}
   0 & I\\
   I & 0
 \end{matrix}}}
\right]$  & Normal ordered expression \\ \hline
$-K_0$    & Anti-normal ordered expression \\ \hline
$\left[{\footnotesize{
\begin{matrix}
   I& 0\\
   0 & I
 \end{matrix}}}
\right]$  & Unit ordered expression \\ 
\hline
General  $K$  & $K$-ordered expression
\end{tabular}
\hfill\parbox[c]{.4\linewidth}
{the product formulas \eqref{eq:KK} give the Weyl ordered 
expression, the normal ordered expression, 
the antinormal ordered expression respectively, but 
the last one, called the unit ordered expression 
is not so familiar in physics.

For each ordered expression, the  product formulas are given  
respectively by the following formulas:}
\begin{equation}\label{ppformula}
 \begin{aligned}[c]
f({\pmb u}){*{_{_0}}}g({\pmb u})=&
f\exp 
\frac{\h i}{2}\{\overleftarrow{\partial_{v}} 
     {\wedge}\overrightarrow{\partial_{u}}\}g,
     \quad{\text{(Moyal product formula)}}\\
f({\pmb u}){*{_{_{K_0}}}}g({\pmb u})=&
f\exp {\h i}\{\overleftarrow{\partial_{v}}\,\, 
       \overrightarrow{\partial_{u}}\}g,
       \qquad{\text{($\Psi$DO-product formula)}}  \\
f({\pmb u}){*{_{_{{-}K_0}}}}g({\pmb u})=&
f\exp{-\h i}\{\overleftarrow{\partial_{u}}\,\, 
       \overrightarrow{\partial_{v}}\}g,
       \quad{\text{($\overline{\Psi}$DO-product formula)}}   
 \end{aligned}
\end{equation} 
where 
$\overleftarrow{\partial_{v}}{\wedge}
\overrightarrow{\partial_{u}}
=\sum_i(\overleftarrow{\partial_{\tilde{v}_i}}
\overrightarrow{\partial_{\tilde{u}_i}}
-\overleftarrow{\partial_{\tilde{u}_i}}
\overrightarrow{\partial_{\tilde{v}_i}})$ and 
$\overleftarrow{\partial_{v}}\,\, 
       \overrightarrow{\partial_{u}}
=\sum_i\overleftarrow{\partial_{\tilde{v}_i}}\,\, 
       \overrightarrow{\partial_{\tilde{u}_i}}$.

The product formula for the unit ordered expression 
is a bit complicated to write down, 
but it is easy to obtain. For instance, 
$$
u_i{*_{_I}}u_i{=}u_i^2{+}\frac{i\h}{2}, \quad 
u_i{*_{_I}}e^{-\frac{1}{i\h}u_i^2}{=}0{=}
e^{-\frac{1}{i\h}u_i^2}{*_{_I}}u_i 
\quad etc
$$
while the Weyl ordered expression gives 
$$ 
{\tilde v}_i{*_{_0}}e^{-\frac{2}{i\h}{\tilde u}_i{\tilde v}_i}{=}
0{=}e^{-\frac{2}{i\h}{\tilde u}_i{\tilde v}_i}{*_{_0}}{\tilde u}_i. 
$$ 
Formulas of unit ordered expression mainly obtained via the 
intertwiners mentioned above. 

\bigskip
For ${\pmb a}, {\pmb b}\in{\mathbb C}^{2m}$, we set 
$\langle{\pmb a}{\varGamma},{\pmb b}\rangle
=\sum_{ij=1}^{2m}{\varGamma}_{ij}a_ib_j$, 
$\langle{\pmb a},{\pmb u}\rangle=\sum_{i=1}^{2m} a_iu_i$. These 
will be denoted also by 
${\pmb a}{\varGamma}\,{}^t{\pmb b}$ and 
$\langle{\pmb a},{\pmb u}\rangle={\pmb a}\,{}^t\!{\pmb u}$. 

\bigskip 
For $f(\pmb u)\in H{\!o}l({\mathbb C}^{2m})$, 
the direct calculation via the product formula 
\eqref{eq:KK} by using Taylor expansion gives the following:

\begin{equation}\label{extend}
\begin{aligned}
&e^{s\frac{1}{i\h}\langle{\pmb a},{\pmb u}\rangle}
{*_{_K}}f({\pmb u})=
e^{s\frac{1}{i\h}\langle{\pmb a},{\pmb u}\rangle}
f({\pmb u}{+}\frac{s}{2}{\pmb a}(K{+}J)),\\ 
&
f({\pmb u}){*_{_K}}
e^{-s\frac{1}{i\h}\langle{\pmb a},{\pmb u}\rangle}=
f({\pmb u}{+}\frac{s}{2}{\pmb a}(-K{+}J))
e^{-s\frac{1}{i\h}\langle{\pmb a},{\pmb u}\rangle}
\end{aligned} 
\end{equation} 
as natural extension of the product formula. This gives also the 
associativity of computations involving two functions 
of exponential growth and a holomorphic function. 

\medskip
Throughout this series, we use notation ${:}{\bullet}{:}_{_K}$ to indicate
the expression parameter for elements of $W_{\h}(2m)$. For instance, we write 
$$
{:}u_i{*}u_j{:}_{_K}{=}u_iu_j{+}\frac{i\h}{2}(K{+}J)_{ij}, \quad  etc.
$$

\subsubsection{For the case $m=1$} 
In the case $m=1$, it is convenient to use notations 
$(\tilde u,\tilde v)$ for $(u_1, u_2)$.
A remarkable feature of the first three formulas of
\eqref{ppformula} is seen as follows:
$$
\begin{aligned}
&{:}\sum a_{kl}{\tilde u}_*^k{*}{\tilde v}_*^l{:}_{_{K_0}}=\sum a_{kl}\tilde u^k\tilde v^l, \quad
(\text{normal ordering}), \\
&{:}\sum a_{kl}\tilde v_*^k{*}\tilde u_*^l{:}_{_{{-}K_0}}=\sum a_{kl}\tilde v^k\tilde u^l,\quad 
(\text{anti-normal ordering}),\\
&{:}\frac{1}{2}(\tilde u{*}\tilde v{+}\tilde v{*}\tilde u){:}_0=\tilde u\tilde v, \quad 
 {:}\frac{1}{3}(\tilde u^2{*}\tilde v{+}
\tilde u{*}\tilde v{*}\tilde u{+}\tilde v^2{*}\tilde u){:}_0
=\tilde u^2\tilde v,\\
&{:}\frac{1}{6}(\tilde u^2{*}\tilde v^2{+}
\tilde u{*}\tilde v{*}\tilde u{*}\tilde v{+}
\tilde u{*}\tilde v^2{*}\tilde u{+}
\tilde v^2{*}\tilde u^2{+}\tilde v{*}\tilde u{*}\tilde v{*}\tilde u{+}
\tilde v{*}\tilde u^2{*}\tilde v){:}_0=\tilde u^2\tilde v^2,\quad etc.. 
\end{aligned}
$$
In general, define  $W_*(\tilde u^k\tilde v^l)$ by $\frac{1}{(k{+}l)!}\sum X_1{*}X_2{*}\cdots{*}X_{k{+}l}$,
where $X_i$ is $\tilde u$ or $\tilde v$ and the summation runs through all possible rearrangement of
$\tilde u^k{*}\tilde v^l$. 
\begin{equation}\label{nikou}
(\tilde u{+}\tilde v)_*^n=\sum_k \,\,{}_nC_kW_*(\tilde u^k\tilde v^{n{-}k}).
\end{equation}
It is easy to see 
$$
{:}\sum a_{kl}W_*(\tilde u^k\tilde v^l){:}_{0}=\sum a_{kl}\tilde u^k\tilde v^l.
$$ 

The next result is trivial, but important.
\begin{prop}\label{trivrmk} 
If $K$ is fixed, then 
every entire function $f(\tilde u,\tilde v)=\sum a_{kl}\tilde u^k\tilde v^l$ can be viewed as 
a $K$-ordered expression of an element.  
\end{prop}
However it is not easy to write down the relations between 
elements written by different expression parameters. 

\medskip
Set 
$H_*=a\tilde u_*^2{+}b\tilde v_*^2{+}2c\tilde u{\ctt}\tilde v$, 
$\tilde u{\ctt}\tilde v=\frac{1}{2}(\tilde u{*}\tilde v{+}\tilde v{*}\tilde u)$, and  
$c^2{-}ab=D$. It is easy to see that 
${:}H_*{:}_0=a\tilde u^2{+}b\tilde v^2{+}2c\tilde u\tilde v$.
The normal ordered expression of 
the $*$-exponential function 
$e_*^{t(a\tilde u^2+b\tilde v^2+2c\tilde u{\ctt}\tilde v)}$ 
is given by \eqref{eq:4.14} by removing
$e^{-itc}$ in the amplitude part.
In this section, its Weyl ordered expression 
${:}e_*^{t(a\tilde u^2+b\tilde v^2+2c\tilde u{\ctt}\tilde v)}{:}_{0}$ will be given. 

\medskip
For that purpose we set 
${:}e_*^{t(a\tilde u^2+b\tilde v^2+2c\tilde u{\ctt}\tilde v)}{:}_0= F(t,\tilde u,\tilde v)$ 
and consider the real analytic solution of the evolution equation 
\begin{equation}
  \label{eq:siki}
\frac{\partial}{\partial t}F(t,\tilde u,\tilde v)=
(a\tilde u^2\!+\!b\tilde v^2\!+\!2c\tilde u\tilde v){*_0}F(t,\tilde u,\tilde v), \quad 
F(0,\tilde u,\tilde v)=1
\end{equation}
By the Moyal product formula in \S\,\ref{formula00}, we have 
$$
\begin{aligned}
(a\tilde u^2\!+&\!b\tilde v^2\!+\!2c\tilde u\tilde v){*_0}F(t,\tilde u,\tilde v)\\
=&(a\tilde u^2\!+\!b\tilde v^2\!+\!2c\tilde u\tilde v)F 
+{\hbar i}\{(b\tilde v\!+\!c\tilde u)\partial_{\tilde u}F-(a\tilde u\!+\!c\tilde v)\partial_{\tilde v}F\}\\ 
{}&-\frac{\hbar^2}{4}\{b\partial_{\tilde u}^2F\!-\!
2c\partial_{\tilde v}\partial_{\tilde u}F\!+\!a\partial_{\tilde v}^2F\}
\end{aligned}
$$
Keeping the uniqueness of the real analytic solution in mind, we set
by using a function $f(x)$ of one variable   
$$
{:}e_*^{t(a\tilde u^2+b\tilde v^2+2c\tilde u\tilde v)}{:}_0= 
f_t(a\tilde u^2{+}b\tilde v^2{+}2c\tilde u\tilde v)
$$ 
to obtain a simplified form
$$
\begin{aligned}
(a\tilde u^2{+}b\tilde v^2{+}2c\tilde u\tilde v)&{*_0}f_t(a\tilde u^2{+}b\tilde v^2{+}2c\tilde u\tilde v)\\
=(a\tilde u^2{+}&b\tilde v^2{+}2c\tilde u\tilde v)f_t(a\tilde u^2{+}b\tilde v^2{+}2c\tilde u\tilde v)\\
   -{\hbar^2}&(ab{-}c^2)(f'_t(a\tilde u^2{+}b\tilde v^2{-}2c\tilde u\tilde v)\\
     +&f''_t(a\tilde u^2{+}b\tilde v^2{+}2c\tilde u\tilde v)(a\tilde u^2{+}b\tilde v^2{+}2c\tilde u\tilde v))
\end{aligned}
$$
Setting $x=a\tilde u^2+b\tilde v^2+2c\tilde u\tilde v$, we obtain the equation
\begin{equation}\label{eq:diff-eq-sq}
\frac{d}{dt}f_t(x) =xf_t(x)+{\hbar^2}D(f'_t(x)+xf''_t(x))
\end{equation}
where $D=c^2-ab$ is the discriminant of $H_*$.

\begin{lem}
  \label{kai}
The solution of the differential equation 
\eqref{eq:diff-eq-sq} with the initial function $1$ is  
$$
{:}e_*^{tH_*}{:}_0=\frac{1}{\cos(\hbar{\sqrt{D}}t)}
     \exp\{{:}H_*{:}_0
\frac{1}{\hbar\sqrt{D}}\tan(\hbar{\sqrt{D}}\,\,t)\}. 
$$
\end{lem}

\noindent 
{\bf Proof}\,\,Set 
$f_t(x)=g(t)e^{h(t)x}$. Plugging this to obtain  
$$
\big\{g'(t)-{D\hbar^2}g(t)h(t)+ 
 xg(t)\{h'(t)-1-{D\hbar^2}h(t)^2\}\big\}e^{h(t)x}=0.
$$
Hence, $h'(t)-1-{D\hbar^2}h(t)^2=0$. By this 
$h(t)$ is obtained as 
$$
h(t)=\frac{1}{\hbar\sqrt{D}}\tan(\hbar(\sqrt{D})t).
$$ 
The sign ambiguity of $\sqrt{D}$ does not suffer the 
result.
 
Next, solving  
$$
g'(t)-g(t)D\hbar^2 
   \frac{1}{\hbar\sqrt{D}}\tan(\hbar(\sqrt{D})t)=0
$$
we have 
$g(t)= \frac{1}{\cos(\hbar(\sqrt{D})t)}$.
The sign ambiguity of $\pm\sqrt{D}$ does not suffer the 
result, and $t$ is allowed to be a complex number.
${}$\hfill$\Box$

\medskip
Consequently, we have the following:
\begin{thm}\label{sisuu}
The Weyl ordered expression of the 
$*$-exponential function 
$e_*^{t(a\tilde u^2{+}b\tilde v^2{+}2c\tilde u{\ctt}\tilde v)}$ is given by 
$$
{:}e_*^{t(a\tilde u^2{+}b\tilde v^2{+}2c\tilde u{\ctt}\tilde v)}{:}_0=
\frac{1}{\cos(\hbar{\sqrt{D}\,t})}
 \exp\left(\frac{1}{\hbar\sqrt{D}}
\tan(\hbar{\sqrt{D}\,t})(a\tilde u^2{+}b\tilde v^2{+}2c\tilde u\tilde v)\right) 
$$
where  
$\frac{1}{\hbar\sqrt{D}}\tan(\hbar{\sqrt{D}\,t}){=}t$ 
in the case $D{=}0$. 

\noindent
${:}e_*^{t(a\tilde u^2{+}b\tilde v^2{+}2c\tilde u{\ctt}\tilde v)}{:}_0$ is singular at 
$t{=}\frac{1}{\sqrt{D}}\frac{\pi}{2\h}(2k{+}1)$, 
$k{\in}\mathbb Z$,  and \\ 
${:}e_*^{t(a\tilde u^2{+}b\tilde v^2{+}2c\tilde u{\ctt}\tilde v)}{:}_0{=}{-}(-1)^k$ 
if $t{=}\frac{1}{\sqrt{D}}\frac{\pi}{\h}(2k{+}1)$, 
$k{\in}\mathbb Z$.
\end{thm}

\noindent
{\bf Remark}.\, A partial polar element ${\e}_{00}(\pmb a)$ cannot be
expressed by the Weyl ordered expression.  

\bigskip
If $D\not=0$, then this case is represented by the case 
$D=1$ where $H_*$ is viewed as  
$$
H_*=\frac{1}{2}\big(
(\alpha \tilde u{+}\beta \tilde v){*}(\gamma \tilde u{+}\delta \tilde v){+}
(\gamma \tilde u{+}\delta \tilde v){*}
(\alpha \tilde u{+}\beta \tilde v)
\big), \quad 
[(\alpha \tilde u{+}\beta \tilde v),(\gamma \tilde u{+}\delta \tilde v)]=-i\h,
$$
the canonical conjugate pair.

\medskip
If $D=1$, then ${:}e_*^{t(a\tilde u^2{+}b\tilde v^2{+}2c\tilde u{\ctt}\tilde v)}{:}_0$ 
is singular at $t=\pm\frac{\pi}{2\hbar}$ in the 
Weyl ordered expression. However, this does not imply that  
$e_*^{t(a\tilde u^2{+}b\tilde v^2{+}2c\tilde u{\ctt}\tilde v)}$ is singular at 
$t{=}\pm\frac{\pi}{2\hbar}$, since 
this singular point disappears in the normal ordered 
expression as it will be seen in 
\S\,\,\ref{anomalousexp}. Singular points depend on expression
parameters. 
%

\bigskip 
By noting 
$\cosh(is)=\cos s$, $\tanh(is)=i\tan s$, we see also  
\begin{equation}
  \label{eq:koki0}
{:}e_*^{t(a\tilde u^2+b\tilde v^2+2c\tilde u{\ctt}\tilde v)}{:}_0=
\frac{1}{\cosh(\hbar{\sqrt{-D}}\,\,t)}
 e^{(a\tilde u^2+b\tilde v^2+2c\tilde u\tilde v)
\frac{1}{\hbar\sqrt{-D}}\tanh(\hbar{\sqrt{-D}}\,t)} 
\end{equation}

\medskip
Replacing $t$ by $t/\h$ we see that 
\begin{equation}
{:}e_*^{it\frac{1}{i\h}H_*}{:}_0=
\frac{1}{\cos{\sqrt{D}}t}
e^{\frac{i}{i\h\sqrt{D}}(\tan{\sqrt{D}}t){:}H_*{:}_0}
\end{equation}
It may be better to rewrite this as 
\begin{equation}
{:}e_*^{t\frac{1}{i\h}H_*}{:}_0=
\frac{1}{\cosh{\sqrt{D}}t}
e^{\frac{1}{i\h\sqrt{D}}(\tanh{\sqrt{D}}t){:}H_*{:}_0}.
\end{equation}
This is rapidly decreasing on ${\mathbb R}$. 

Although ${:}e_*^{t\frac{1}{i\h}H_*}{:}_0$ has  singularities, the exponential law holds 
by the uniqueness of real analytic solution of the defining equation:
\begin{equation}\label{eq:exprule}
  \begin{aligned}
e_*^{s(a\tilde u^2+b\tilde v^2+2c\tilde u{\ctt}\tilde v)}&*
e_*^{t(a\tilde u^2+b\tilde v^2+2c\tilde u{\ctt}\tilde v)}=
e_*^{(s+t)(a\tilde u^2+b\tilde v^2+2c\tilde u{\ctt}\tilde v)}\\
e_*^{t(a\tilde u^2+b\tilde v^2+2c\tilde u{\ctt}\tilde v+\alpha)}&
=e^{\alpha t}e_*^{t(a\tilde u^2+b\tilde v^2+2c\tilde u{\ctt}\tilde v)},\quad 
\alpha\in{\mathbb C}.     
\end{aligned}
\end{equation}

\medskip

Recalling $2\tilde u{*}\tilde v=2\tilde u{\ctt}\tilde v{-}i\h$, we have by 
\eqref{eq:exprule}
\begin{equation}\label{vacuua}
\begin{aligned}
&\lim_{t\to\infty}
{:}e_*^{t\frac{1}{i\h}(a\tilde u^2+b\tilde v^2+2c\tilde u{*}\tilde v)}{:}_0
=\lim_{t\to\infty}\frac{e^{t}}{\cosh t}
e^{\frac{1}{i\h}(\tanh t)(a\tilde u^2{+}b\tilde v^2{+}2c\tilde u\tilde v)}
=e^{\frac{1}{i\h}(a\tilde u^2{+}b\tilde v^2{+}2c\tilde u\tilde v)},\\
&\lim_{t\to -\infty}
{:}e_*^{t\frac{1}{i\h}(a\tilde u^2+b\tilde v^2+2c\tilde v{*}\tilde u)}{:}_0
=\lim_{t\to -\infty}\frac{e^{-t}}{\cosh t}
e^{\frac{1}{i\h}(\tanh t)(a\tilde u^2{+}b\tilde v^2{+}2c\tilde u\tilde v)}
=e^{-\frac{1}{i\h}(a\tilde u^2{+}b\tilde v^2{+}2c\tilde u\tilde v)}.
\end{aligned}
\end{equation}
These have idempotent property by the exponential law. 
These elements  are  called {\bf vacuums} 
in what follows. Note that such elements are not 
classical elements for they are not defined for 
$\h=0$. On the other hand, we easily see  
$$
\lim_{t\to -\infty}
{:}e_*^{t\frac{1}{i\h}(a\tilde u^2+b\tilde v^2+2c\tilde u{*}\tilde v)}{:}_0
=0, \quad 
\lim_{t\to \infty}
{:}e_*^{t\frac{1}{i\h}(a\tilde u^2+b\tilde v^2+2c\tilde v{*}\tilde u)}{:}_0
= 0.
$$
Therefor an operation such as 
$$
\lim_{t\to \infty}
e_*^{t\frac{1}{i\h}(a\tilde u^2+b\tilde v^2+2c\tilde u{*}\tilde v)}{*}p(\tilde u,\tilde v){*}
e_*^{-t\frac{1}{i\h}(a\tilde u^2+b\tilde v^2+2c\tilde u{*}\tilde v)}
$$
is not defined.

\subsection{Star-exponential functions of linear functions}

By a direct calculation of intertwiners, we see that     
\begin{equation}
  \label{eq:intwin}
I_{_K}^{^{K'}}(e^{\frac{1}{i\h}\langle{\pmb a},{\pmb u}\rangle})
=e^{\frac{1}{4i\h}\langle{\pmb a}(K'{-}K),{\pmb a}\rangle}
e^{\frac{1}{i\h}\langle{\pmb a},{\pmb u}\rangle}.   
\end{equation}
Hence, $\{e^{\frac{1}{4i\h}\langle{\pmb a}K,{\pmb a}\rangle}
e^{\frac{1}{i\h}\langle{\pmb a},{\pmb u}\rangle}; 
K\in{\mathfrak S}_{\mathbb C}(2m)\}$ is a parallel section of 
$\coprod_{K\in{\mathfrak S}_{\mathbb C}(2m)}
{H{\!o}l}({\Bbb C}^{2m}).$

We shall denote this element symbolically by 
$e_*^{\frac{1}{i\h}\langle{\pmb a},{\pmb u}\rangle}$. 
Namely, we define   
\begin{equation}
  \label{eq:tempexp}
:e_*^{\frac{1}{i\h}\langle{\pmb a},{\pmb u}\rangle}:_{_K}
=e^{\frac{1}{4i\h}\langle{\pmb a}K,{\pmb a}\rangle}
e^{\frac{1}{i\h}\langle{\pmb a},{\pmb u}\rangle}
=e^{\frac{1}{4i\h}\langle{\pmb a}K,{\pmb a}\rangle
{+}\frac{1}{i\h}\langle{\pmb a},{\pmb u}\rangle}.  
\end{equation}

It is remarkable that if $K=0$, then 
$:e_*^{\frac{1}{i\h}\langle{\pmb a},{\pmb u}\rangle}:_{_K}=
e^{\frac{1}{i\h}\langle{\pmb a},{\pmb u}\rangle}$, that is,
$*$-exponential functions of linear functions are 
ordinary exponential functions.

By using the product formula for $K$-ordered expression, 
we have easily the exponential law  
$$
{:}e_*^{s\frac{1}{i\h}\langle{\pmb a},{\pmb u}\rangle}
{:}_{_{K}}{*_{_{K}}}
{:}e_*^{t\frac{1}{i\h}\langle{\pmb a},{\pmb u}\rangle}
{:}_{_{K}}=
{:}e_*^{(s{+}t)\frac{1}{i\h}\langle{\pmb a},{\pmb u}\rangle}
{:}_{_{K}},\,\,
\forall K\in {\mathfrak S}(2m).
$$
The exponential law may be written by omitting the suffix $K$ as  
$$
e_*^{s\frac{1}{i\h}\langle{\pmb a},{\pmb u}\rangle}{*}
e_*^{t\frac{1}{i\h}\langle{\pmb a},{\pmb u}\rangle}=
e_*^{(s{+}t)\frac{1}{i\h}\langle{\pmb a},{\pmb u}\rangle},\quad 
e^{s}e_*^{t\frac{1}{i\h}\langle{\pmb a},{\pmb u}\rangle}=
e_*^{s{+}t\frac{1}{i\h}\langle{\pmb a},{\pmb u}\rangle}
$$
together with the exponential law with the 
ordinary exponential functions.

\bigskip

Furthermore, for every $K$, 
$e_*^{\frac{s}{i\h}\langle{\pmb a},{\pmb u}\rangle}$ 
is the solution of the evolution equation 
$$
\frac{d}{dt}{:}e_*^{\frac{s}{i\h}\langle{\pmb a},{\pmb u}\rangle}
{:}_{_{K}}
=\frac{1}{i\h}{:}\langle{\pmb a},{\pmb u}\rangle{:}_{_{K}}{*_{_K}}
{:}e_*^{\frac{s}{i\h}\langle{\pmb a},{\pmb u}\rangle}{:}_{_{K}}
\,\,{\text{with initial data}}\,\, {:}1{:}_{_{K}}=1.
$$
Note that ${:}\langle{\pmb a},{\pmb u}\rangle{:}_{_{K}}
=\langle{\pmb a},{\pmb u}\rangle$. 
$e_*^{s\frac{1}{i\h}\langle{\pmb a},{\pmb u}\rangle}=
\{e^{s^2\frac{1}{4i\h}\langle{\pmb a}K\,{\pmb a}\rangle}
e^{s\frac{1}{i\h}\langle{\pmb a},{\pmb u}\rangle}; 
K\in{\mathfrak S}(2m)\}$ forms a one parameter 
group of parallel sections. 

\bigskip
By applying \eqref{extend} to 
${:}e_*^{\pm s\frac{1}{i\h}\langle{\pmb a},{\pmb u}\rangle}{:}_{_K}$
carefully, we have for every $f\in H\!ol({\mathbb C}^n)$ that  
\begin{equation}\label{adjointlike}
{:}(e_*^{s\frac{1}{i\h}
\langle{\pmb a},{\pmb u}\rangle}{*}f_*({\pmb u}))
{*}e_*^{-s\frac{1}{i\h}\langle{\pmb a},{\pmb u}\rangle}{:}_{_K}
={:}f_*({\pmb u}{+}{s}{\pmb a}J){:}_{_K}=
{:}e_*^{s\frac{1}{i\h}
\langle{\pmb a},{\pmb u}\rangle}{*}(f_*({\pmb u})
{*}e_*^{-s\frac{1}{i\h}\langle{\pmb a},{\pmb u}\rangle}){:}_{_K}.
\end{equation}
This gives also the associativity and the real analyticity of 
$e_*^{s\frac{1}{i\h}
\langle{\pmb a},{\pmb u}\rangle}{*}f_*({\pmb u})
{*}e_*^{-s\frac{1}{i\h}\langle{\pmb a},{\pmb u}\rangle}$ in $s$.
However if we know the associativity in advance by using 
Theorem\,\ref{assocthm} for instance, then 
it is better to compute as follows: 

Differentiating  
$F_*(s)= e_*^{s\frac{1}{i\h}
\langle{\pmb a},{\pmb u}\rangle}{*}f_*({\pmb u})
{*}e_*^{-s\frac{1}{i\h}\langle{\pmb a},{\pmb u}\rangle}$ in $s$, we
have 
$$
\frac{d}{ds}F_*(s)=
[\frac{1}{i\h}\langle{\pmb a},{\pmb u}\rangle, F_*(s)],\quad 
F_*(0)=f_*({\pmb u}). 
$$
On the other hand, $f_*({\pmb u}{+}{s}{\pmb a}J)$ satisfies the same
equation 
$$
\frac{d}{ds}f_*({\pmb u}{+}{s}{\pmb a}J)=
[\frac{1}{i\h}\langle{\pmb a},{\pmb u}\rangle,f_*({\pmb u}{+}{s}{\pmb a}J)].
$$
Thus, the uniqueness of real analytic solution gives \eqref{adjointlike}.

\bigskip

The product formula gives
\begin{equation}\label{prodformula}
{:}e_*^{\frac{1}{i\h}\langle{\pmb a},{\pmb u}\rangle}{*}
 e_*^{\frac{1}{i\h}\langle{\pmb b},{\pmb u}\rangle}{:}_{_K}
=e^{\frac{1}{2i\h}{\langle\pmb a}{J},{\pmb b}\rangle}
{:}e_*^{\frac{1}{i\h}\langle({\pmb a}{+}{\pmb b}),
{\pmb u}\rangle}{:}_{_K}.
\end{equation}
This is equivalent with 
${:}e_*^{\langle{\pmb a},{\pmb u}\rangle}{*}
 e_*^{\langle{\pmb b},{\pmb u}\rangle}{:}_{_K}
=e^{\frac{i\h}{2}{\langle\pmb a}{J},{\pmb b}\rangle}
{:}e_*^{\langle({\pmb a}{+}{\pmb b}),
{\pmb u}\rangle}{:}_{_K}$.
This forms a noncommutative group isomorphic to the group  
${\mathbb C}^{2m}\times {\mathbb C}$ with the 
group structure  
\begin{equation}\label{Heisenberg}
(\pmb a,\lambda)*(\pmb b,\mu)=
(\pmb a{+}\pmb b, \lambda{+}\mu{+}
\frac{1}{2}{\langle{\pmb a}{J},{\pmb b}\rangle}). 
\end{equation}
This is viewed as a central extension of the abelian group 
${\mathbb C}^{2m}$, called sometimes Heisenberg group.
The algebra generated by 
$\{e_*^{\frac{1}{i\h}\langle{\pmb a},{\pmb u}\rangle}; {\pmb a}\in
{\mathbb C}^{n}\}$ is called the {\bf noncommutative torus}.

\bigskip
\noindent
{\bf Remark for notations of $*$-products}.\,\,
Since $*$-commutators are independent of expression parameters, 
we often omit the suffix ${:}\,\,{:}_{_K}$ or $*_{_K}$ in 
 computations involving only commutation relations.

It is obvious that the correspondence 
$$
{\pmb x}\to e_*^{\langle{\pmb x},{\pmb u}\rangle},\quad c\to e^{c}
$$
gives an isomorphism of Heisenberg group onto the noncommutative torus. 

\begin{center}
\fbox{\parbox[c]{.6\linewidth}{$e_*^{\frac{1}{i\h}\langle{\pmb a},{\pmb u}\rangle}{*}
 e_*^{\frac{1}{i\h}\langle{\pmb b},{\pmb u}\rangle}
=e^{\frac{1}{2i\h}{\langle\pmb a}{J},{\pmb b}\rangle}
e_*^{\frac{1}{i\h}\langle({\pmb a}{+}{\pmb b}),{\pmb u}\rangle}
= e^{\frac{1}{i\h}{\langle\pmb a}J,{\pmb b}\rangle}
 e_*^{\frac{1}{i\h}\langle{\pmb b},{\pmb u}\rangle}{*}
 e_*^{\frac{1}{i\h}\langle{\pmb a},{\pmb u}\rangle}$}}
 \end{center}

\subsubsection{Linear change of generators} 
Next, we consider the effect of a linear change of generators  
$$
{u'}_i=\sum u_kS_i^k, \quad S\in G\!L(n,{\mathbb C}),\quad 
 ({\pmb u}'={\pmb u}S).
$$
By the help that 
$$
\partial_{u_i}=\sum S_i^k\partial_{u'_k},
$$
 the product formula is rewritten 
by using new generators as 
\begin{equation} \label{eq:KK2}
 f*_{_{\Lambda}}g=fe^{\frac{i\h}{2}
(\sum\overleftarrow{\partial_{u'_i}}
({}^t\!S{\Lambda}S)^{ij}\overrightarrow{\partial_{u'_j}})}g.
\end{equation}
Thus the notation $*_{_\Lambda}$ is better to be replaced by 
$*_{_{\Lambda'}}$ where $\Lambda'{=}{}^t\!S{\Lambda}S$. 
Therefor the algebraic structure of 
$({\mathbb C}[\pmb u], *_{_\Lambda})$ 
depends only on the conjugacy class of the skew part $J$.

\medskip
If ${}^t\!SJS=J$, that is, $S$ is a symplectic linear change 
of generators 
$$
{u'}_i=\sum u_kS_i^k, \quad S\in Sp(m,{\mathbb C}), 
$$
 then the mapping ${\pmb u}\to {\pmb u}'$ 
does not change the algebraic structure. 
Thus, a symplectic change of generators is recovered by    
the intertwiner $I_{_K}^{^{{}^t\!SKS}}$.  
Change of generators are viewed often as coordinate transformations, 
but note here that $I_{_K}^{^{{}^t\!SKS}}$ 
is something like the ``square root'' of 
symplectic  coordinate transformations.  

\medskip
Since $\det S=1$ for $S\in Sp(m,{\mathbb C})$, we see 
$\det{}^t\!SKS=\det K$, hence the isomorphic 
change by the intertwiner $I_{_K}^{^{K'}}$ cannot be  
recovered by a coordinate transformation if 
$\det K\not=\det K'$. 

\medskip
Even if $S\in G\!L(n,{\mathbb C})$ is not in $Sp(m,{\mathbb C})$, 
setting ${u'}_i=\sum u_kS_i^k $ and  $J'={}^t\!SJS$ gives an 
isomorphism 
$$
\Phi_{_S}: ({\mathbb C}[\pmb u];*_{J})\to
({\mathbb C}[\pmb u'];*_{J'}).
$$ 
of Weyl algebras.

Keeping that ${\pmb u}$ are complex variables, we 
have the following formula: 
Let ${\pmb u}={\pmb u}'S{+}{\pmb b}$, 
$S\in Sp(m,{\mathbb C})$. Then, 
$$
e_*^{\frac{1}{i\h}\langle{\pmb a},{\pmb u}'S{+}{\pmb b}\rangle}=
e^{\frac{1}{i\h}\langle{\pmb a},{\pmb b}\rangle}
e_*^{\frac{1}{i\h}\langle{\pmb a}{}^{t}\!S,{\pmb u}'\rangle}
$$
and 
$$ 
{:}e_*^{\frac{1}{i\h}\langle{\pmb a},{\pmb u}'S{+}{\pmb b}\rangle}{:}_{_{K,{\pmb u}'}}
=e^{\frac{1}{4i\h}\langle{\pmb a}{}^{t}\!S\!K\!S,{\pmb a}\rangle
{+}\frac{1}{i\h}\langle{\pmb a},{\pmb u}'S{+}{\pmb b}\rangle}
=e^{\frac{1}{4i\h}\langle{\pmb a}{}^{t}\!S\!K\!S,{\pmb a}\rangle
{+}\frac{1}{i\h}\langle{\pmb a},{\pmb u}\rangle}
$$
where ${:}{\,\,\,}{:}_{_{K,{\pmb u}'}}$ means the 
$K$-ordered expression with respect
to the generator system ${\pmb u}'$. The above formula may be 
written as the formula
\begin{equation}
\label{eq:exchform}
{:}e_*^{\frac{1}{i\h}
\langle{\pmb a},{\pmb u}\rangle}{:}_{_{K,{\pmb u}'}}  
=
{:}e_*^{\frac{1}{i\h}
\langle{\pmb a},{\pmb u}\rangle}{:}_{_{{}^{t}\!S\!K\!S,{\pmb u}}}.
\end{equation}
By this formula, we see that the change of expression parameters  
can be traced by the change of generator systems. 

\subsection{Remarks on real analyticity and associativity}

A mapping $f: U\to F$ from an open subset $U$ of $\mathbb R$ 
into a Fr{\'e}chet space $F$ is called    
{\bf real analytic}, if for every $a\in U$ there is an $\e(a)>0$
such that $f$ is written in the form  
$$
f(a+s)=\sum_k \frac{1}{k!}a_k s^k, \quad  a_k\in F, \quad 
|s|<\e(a), 
$$
where $a_k$ is given by $a_k=\partial_s^kf|_{s=0}$.

If $F$ is a Banach space and $\sum_k\frac{1}{k!}\|a_k\||s|^k$
converges, then the power series 
$\sum_k a_k s^k$ is said to 
{converge absolutely} under the norm. 

If a Fr{\'e}chet space $F$ is defined by a countable 
family of seminorms  
$\{\|f\|_{\ell}; {\ell}=1,2,3\cdots\}$, then replace this part by 
the absolute convergence of 
$\sum_k\frac{1}{k!}\|a_k\|_{\ell}|s|^k$ 
w.r.t. seminorms $\|\cdot\|_{\ell}$. 
 A power series $\sum_k a_k s^k$ converges if this 
converges absolutely under every seminorms. 

\bigskip
\noindent
{\bf Radius of convergence.}\,\,Suppose a 
Fr{\'e}chet space $F$ is defined by a countable family of 
seminorms $\{\|f\|_{\ell}; {\ell}=1,2,3\cdots\}$.

\begin{lem}
  \label{powerser}
For a power series $\sum_k a_k s^k$, $a_k\in F$,
there exists a unique real number $R$ 
$(0\leq R\leq \infty)$ satisfying $(1)$ 
and $(2)$ below:
\begin{description}
\item[(1)] If $|s|<R$, then the power series 
$\sum_k a_k s^k$ converges 
absolutely under every seminorm $\|\cdot\|_{\ell}$.
 
\item[(2)] If $|s|>R$, then $\sum_k a_k s^k$ does not converge. 
\end{description}
\end{lem}

\medskip
\noindent
{\bf Proof}\,\,\, Suppose $\sum_k a_k s_0^k$ converges  
at $s_0$. Then $a_k s_0^k$ is bounded 
under every seminorm $\|\cdot\|_{\ell}$. Set 
$\sup_k\|a_k s_0^k\|_{\ell}\leq M_{\ell}$. Then for every 
$s$ such that $|s|<|s_0|$ we see 
$$
\sum_k\|a_ks^k\|_{\ell}\leq \sum_k M_{\ell}|s/s_0|^k 
= M_{\ell}\frac{1}{1-|s/s_0|}<\infty.
$$
Then the convergence of $\sum_ka_ks^k$ follows. 
${}$\hfill$\square$

\normalsize
\begin{lem}
  \label{powbibun}
$\sum_{k\geq 0}a_k s^k$ and 
$\sum_{k\geq 1}ka_k s^{k-1}$ have same radius of convergence. 
\end{lem}

Real analyticity is left invariant under every continuous 
linear transformation.
\begin{lem}
  \label{linsbl}
Let $F, G$ be Fr{\'e}chet spaces and 
$\varphi : F\to G$ be a continuous linear mapping.  
If $f: U\to F$ is real analytic, then 
$\varphi f: U\to G$ is also real analytic. 
\end{lem}

\bigskip

\noindent
{\bf Remarks for the associativity}\,\,
Products of exponential functions of quadratic forms 
may not be defined, and even if the product is defined 
associativity may not hold. 
In general, we do not have associativity even for a polynomial $p({\pmb{u}})$
$$
(e^{H({\pmb{u}})}{*}p({\pmb{u}})){*}e^{K({\pmb{u}})},\quad
e^{H({\pmb{u}})}
{*}(p({\pmb{u}}){*}e^{K({\pmb{u}})}), 
$$
since $p({\pmb{u}})$ has two different $*$-inverses in general.

\medskip

However, if we can treat elements in 
$({\mathbb C}[\pmb u][[\h]], {*_{_K}})$, 
the space of formal power series of $\h$,
then $*_{_\Lambda}$-product is always defined by the 
product formula \eqref{eq:KK} and the associativity holds. 

Elements of $H\!ol({\Bbb C}^{n})$ are often 
given as real analytic functions of $\h$ defined on 
certain interval containing $\h=0$.  
The following is easy to see:

\begin{thm}\label{assocthm}
Suppose $f(\h,{\pmb u})$,  
$g(\h,{\pmb u})$ and $h(\h,{\pmb u})$
are given as real analytic function of $\h$ in some interval 
$[0,H]$. If all of these 
$$
f(\h,{\pmb u}){*_{_K}}g(\h,{\pmb u}),\,\, 
(f(\h,{\pmb u}){*_{_K}}g(\h,{\pmb u})){*_{_K}}h(\h,{\pmb u}),\,\, 
g(\h,{\pmb u}){*_{_K}}h(\h,{\pmb u}),\,\, 
f(\h,{\pmb u}){*_{_K}}(g(\h,{\pmb u}){*_{_K}}h(\h,{\pmb u}))
$$
are defined as real analytic functions on $\h\in [0,H]$, then 
the associativity holds: i.e. 
$$
(f(\h,{\pmb u}){*_{_K}}g(\h,{\pmb u})){*_{_K}}h(\h,{\pmb u})
=f(\h,{\pmb u}){*_{_K}}(g(\h,{\pmb u}){*_{_K}}h(\h,{\pmb u})).
$$
\end{thm}
We refer to this theorem as the {\bf formal associativity theorem}.

\noindent 
{\bf Remark 1}. 
In what follows, elements are often given 
in the form $f(\frac{1}{i\h}\varphi(t),{\pmb u})$ 
by using a real analytic function $f(t,{\pmb u})$, $t{\in}[0,T]$, 
where $\varphi(t)$ is a real analytic function such that 
$\varphi(0){=}0$. 
(Cf.\eqref{eq:tempexp}, \eqref{eq:koki0}, \eqref{eq:norexpm}). 
In such a case, 
replacing $t$ by $s\h$ gives a real analytic function 
of $\h$, and such an element is embedded in 
$({\mathbb C}[\pmb u][[\h]], {*_{_K}})$. 
Thus, we can apply the above theorem. We call such elements 
{\bf classical elements}. 
However, there are many elements in 
$H\!ol({\Bbb C}^{n})$ 
written in the form $f(\frac{1}{i\h}\varphi(t),{\pmb u})$ 
such that $\varphi(0){\not=}0$.

Using Lemma\,\ref{frechet}, we have the following:
\begin{lem}\label{realanal}
Let $U$ be an connected open neighborhood of $0$ of 
${\mathbb R}^{\ell}$ 
Suppose $\psi: U\to H{\!o}l({\mathbb C}^n)$ be a
real analytic mapping. Then 
$x\to p({\pmb u}){*}\psi(x){*}q({\pmb u})$ is also a 
real analytic on $U$ for every polynomial 
$p({\pmb u}),\,q({\pmb u})$. 
\end{lem}

\noindent
{\bf Proof}\,\,\, It is easy by using that 
 $X\to p({\pmb u}){*}X{*}q({\pmb u})$ is a continuous 
linear mapping. \hfill $\Box$


In the noncommutative torus multiplicative commutators play the
same role as commutators:
\begin{lem}\label{elementary}
The multiplicative commutator gives
$$
{:}e_*^{-\frac{1}{i\h}\langle{\pmb b},{\pmb u}\rangle}{*}
e_*^{-\frac{1}{i\h}\langle{\pmb a},{\pmb u}\rangle}{*}
e_*^{\frac{1}{i\h}\langle{\pmb b},{\pmb u}\rangle}{*}
e_*^{\frac{1}{i\h}\langle{\pmb a},{\pmb u}\rangle}{:}_{_K}
=e^{\frac{1}{i\h}\langle{\pmb b}J,{\pmb a}\rangle}
$$
which belongs to the center independent of expression parameters.
For the case $m=1$, the multiplicative commutator gives
the area $\langle{\pmb b}J,{\pmb a}\rangle$ 
of the rectangular domain spanned by 
 ${\pmb b}=(b_1,b_2)$ and ${\pmb a}=(a_1,a_2)$.
\end{lem}

On the other hand, regarding $\h$ as a member of generators,  
the Lie algebra generated by $\h$ and 
$\{\langle{\pmb a},{\pmb u}\rangle,\,
{\pmb a}\in {\mathbb C}^{n}\}$ with  
relations $[\tilde u_i,\tilde v_j]=-\sqrt{-1}\h\delta_{ij}$  
is called also the (complex) {\bf Heisenberg Lie algebra}. 
Its universal enveloping algebra is called the {\bf Heisenberg algebra}.
We denote this algebra by ${\mathcal H}(2m)$. 
In contrast with the Weyl algebra  $W_{\h}(2m)$ in previous sections, 
$\h$ is not treated as a scalar, but 
a member of generators, hence $\frac{1}{i\h}$ is not an 
element of ${\mathcal H}(2m)$.

\subsubsection{Subalgebras and their two-sided ideals}

$({H{\!o}l}({\Bbb C}^{2m}),*_{_K})$ contains various 
systems which closed under the $*_{_K}$-product, which 
will be called {\it subalgebras}.  
Weyl algebra $(W_{\h}(2m),*_{_K})$ is a dense subalgebras. 

\begin{lem}\label{Heisenideal}
There is no nontrivial two-sided ideal of the Weyl algebra 
$(W_{\h}(2m),*_{_K})$. On the other hand 
the Heisenberg algebra $({\mathcal H}(2m),*_{_K})$ has 
two-sided ideals corresponding to points of 
${\mathbb C}^{2m}$. 
\end{lem}

\noindent
{\bf Proof}\,\,is easy by observing the following: 
Suppose $\psi$ is a homomorphism of an algebra into 
${\mathbb C}$, and suppose $[x,y]_*=z$, then $\psi(z)=0$. 
It follows that there is no nontrivial two-sided ideal 
of the Weyl algebra $(W_{\h}(2m),*_{_K})$. 

On the other hand
$\h{*}{\mathcal H}(2m)$ is a two-sided ideal of 
${\mathcal H}(2m)$ such that quotient algebra 
is the usual commutative polynomial ring ${\mathbb C}[\pmb u]$. 
It is easy to see that for every 
${\pmb a}\in {\mathbb C}^{2m}$, 
the two-sided ideal of ${\mathbb C}[\pmb u]$ generated by ${\pmb u}{-}{\pmb a}$ is 
pull back to give an nontrivial two-sided ideal of ${\mathcal H}(2m)$. 
\hfill $\Box$

\medskip
Hence, to treat the Heisenberg algebra as a topological algebra, 
it is better to write the generators as 
$\{i\h, \langle{\pmb a},{\pmb u}\rangle,\,{\pmb a}\in {\mathbb C}^{n}\}$ 
without using $\frac{1}{i\h}$. 
Then, the Heisenberg algebra may be treated in 
$(H\!ol({\mathbb C}^{2m+1}),*)$. 
These are seen in \cite{om3}, pp195-200, pp300-305.

\section{Intertwiners for exponential 
functions of quadratic forms}
\label{intwners}

In this section we investigate intertwiners  
on the space of exponential functions of quadratic forms 
${\mathbb C}e^{{\mathfrak S}(2m)}$. This will be used also 
to obtain $K$-ordered expressions of star-exponential 
functions of quadratic forms. 
In the argument in this section, 
the skew part $J$ of $\Lambda=K{+}J$ 
need not be nondegenerate. 
So the arguments can be applied for the case $J=0$.

\subsection{Restrictions to the space of exponential functions}

If the generator system/fundamental coordinate system is fixed,  
infinitesimal intertwiners are viewed naturally a flat connection 
defined on the trivial bundle over the  space of expression parameters. 

Let ${\mathbb C}e^{{\mathfrak S}(2m)}$ be the multiplicative space of all
exponential functions of quadratic forms. We consider 
the product bundle 
$$
\coprod_{K\in{\mathfrak S}(2m)}{\mathbb C}e^{{\mathfrak S}(2m)}
\subset 
\coprod_{K\in{\mathfrak S}(2m)}H\!ol({\mathbb C}^{2m}) 
\quad({\text{parallel subbundle}}).
$$
We restrict the connection (infinitesimal intertwiner) to the subbundle
$$
\coprod_{K\in{\mathfrak S}(2m)}
(e^{{\mathfrak S}(2m)}; {*_{K}})
$$
%
%
A horizontal distribution 
$H_{_K}(ge^{\langle{\pmb u}\frac{1}{i\h}A,{\pmb u}\rangle})$ 
at $ge^{\langle{\pmb u}\frac{1}{i\h}A,{\pmb u}\rangle}$ defined 
 on 
$\coprod_{K\in{\mathfrak S}(2m)}{\mathbb C}e^{{\mathfrak S}(2m)}$
is given by applying the infinitesimal intertwiner $dI_{_K}(K')$ as follows:
$$
H_{_K}(ge^{\langle{\pmb u}\frac{1}{i\h}A,{\pmb u}\rangle})
=\{\big(K'; g(\frac{1}{2}{\rm{Tr}}K'A{+}
  \langle{\pmb u}\frac{1}{i\hbar}AK'A,{\pmb u}\rangle)
  e^{\langle{\pmb u}\frac{1}{i\h}A,{\pmb u}\rangle}\big); 
K'\in{\mathfrak S}(2m)\}.  
$$
The infinitesimal intertwiner/horizontal distribution is viewed as  
a flat connection on these bundles. 
The intertwiners can be viewed as parallel translations,   
though a parallel displacement is not defined on the whole space 
in general. 
However, since functions are restricted  
to the space of exponential functions of quadratic forms, 
the equation of parallel displacement can be solved locally. 

\bigskip

Let ${\pmb u}=(u_1,\dots,u_{2m})$. 
The exact formula to parallel translation is obtained by solving the 
evolution equation 
$$
\frac{d}{dt}g(t)e^{\frac{1}{i\h}
\langle{\pmb u}Q(t),{\pmb u}\rangle}=
\sum_{ij}K^{ij}\partial_{u^i}\partial_{u^j}
\big(g(t)e^{\frac{1}{i\h}
\langle{\pmb u}Q(t),{\pmb u}\rangle}\big), 
\quad Q(0)=A, \quad g(0)=g 
$$
by setting 
\begin{equation}
  \label{eq:intmultisol}
e^{t\sum_{ij}K^{ij}\partial_{u^i}\partial_{u^j}}
(ge^{\frac{1}{i\h}\langle{\pmb u}A,{\pmb u}\rangle}) 
=g(t)e^{\frac{1}{i\h}\langle{\pmb u}Q(t),{\pmb u}\rangle}. 
\end{equation}
A direct calculation gives   
$$
\sum_iK^{ij}\partial_{u^i}\partial_{u^j}
\big(g(t)e^{\frac{1}{i\h}
\langle{\pmb u}Q(t),{\pmb u}\rangle}\big)
=g(t)\Big(2{\rm{Tr}}K\frac{1}{i\hbar}Q(t)
+4\frac{1}{(i\hbar)^2}(QKQ)_{ij}u^iu^j\Big)
e^{\frac{1}{i\h}\langle{\pmb u}Q(t),{\pmb u}\rangle}.
$$
By uniqueness of the real analytic solution, 
we only have to solve a system of 
ordinary differential equations: 
$$
\left\{
  \begin{aligned}[c]
\frac{d}{dt}Q(t)&=\frac{4}{i\hbar}Q(t)KQ(t)\\
\frac{d}{dt}g(t)&=g(t)(\frac{2}{i\hbar}{\rm{Tr}}KQ(t))    
  \end{aligned}
\right. \qquad Q(0)=A, \quad g(0)=g. 
$$ 
Hence, we have $Q(t)=\frac{1}{I-\frac{4t}{i\hbar}AK}A$,\,\,  
$g(t)=g(\det(I-\frac{4t}{i\hbar}AK))^{-1/2}$. 

Here, the inverse matrix
of $X$ is denoted by $\frac{1}{X}$. Note that 
$\frac{1}{X}\frac{1}{Y}=\frac{1}{YX}$.
It is easy to check that $\frac{1}{I{-}AK}A$ is a symmetric 
matrix by the
bumping identity 
\begin{equation}
  \label{eq:bump}
\frac{1}{I{-}AK}A=A\frac{1}{I{-}KA}.  
\end{equation}
Setting $t=\frac{\h i}{4}$, we have the intertwiner 
$I_{0}^{^{K}}$: 
\begin{equation}
  \label{eq:intwnr}
Q(\frac{\h i}{4})=\frac{1}{I-AK}A, \quad 
g(\frac{\h i}{4})=g(\det(I-AK))^{-\frac{1}{2}}.
\end{equation} 
For simplicity, we denote 
$ge^{\frac{1}{i\hbar}\langle{\pmb u}A,{\pmb u}\rangle}$ 
by $(g; A)$, and 
we call $g$ and $A$ the {\it amplitude} and the {\it phase} 
part of $ge^{\frac{1}{i\hbar}\langle{\pmb u}A,{\pmb u}\rangle}$.
In this notation, we see that 
$$
I_{0}^{^K}(g; A)=\Big(g\det(I-AK)^{-\frac{1}{2}}; T_K(A)\Big),
$$
where 
$T_{K}: {\mathfrak S}(2m)\to{\mathfrak S}(2m),\quad 
T_{K}(A)=\frac{1}{I{-}AK}A$ 
is viewed as the phase part of the intertwiner $I_0^{^K}$. 

\medskip
Computing the inverse $I^{0}_{_K}=(I_{0}^{^K})^{-1}$, and 
the composition $I_{0}^{^{K'}}I^{0}_{_K}$, we easily see 
\begin{equation}\label{inter}
I_{_{K}}^{^{K'}}(g;A) 
=\Big(g\det(I{-}A(K'{-}K))^{-\frac{1}{2}};
\frac{1}{I{-}A(K'{-}K)}A\Big).
\end{equation}
This mapping is singular at $A$ such that 
$\det(I{-}A(K'{-}K)){=}0$, and the sign ambiguity 
cannot be removed. 
$T_{_K}^{^{K'}}(A){=}\frac{1}{I{-}A(K'{-}K)}A$ is viewed 
as the phase part of the intertwiner. 

Note that the identities  
$$
T_{_K}^{^{K'}}{\sim}T_{K'}(T_{K})^{-1}, 
\quad 
I_{_K}^{^{K'}}{\sim}I_0^{^{K'}}I_{_K}^0
$$
hold. Here $\sim$ means the equality in algebraic calculations 
such as $x/x{=}1$, $\sqrt{1{+}x}/\sqrt{1{+}x}{=}1$.  
Singularities are moved by this algebraic trick. 
Moving branched singularities are the remarkable 
feature of this calculus.  

By the  concrete form of intertwiners for exponential 
functions of quadratic forms, we see the following 
%
%
\begin{thm}
There is no globally defined parallel section of 
$\coprod_{K\in{\mathfrak S}(2m)}
{\mathbb C}e^{{\mathfrak S}(2m)}$
except constant scalar sections $($trivial sections$)$, 
and every nontrivial parallel section is two-valued. 
\end{thm}

Since setting ${}^t\!\tilde{\pmb a}\tilde{\pmb a}=(a_ia_j)=A$ we see 
$$
{:}\langle\tilde{\pmb a},\tilde{\pmb u}\rangle{*}
\langle\tilde{\pmb a},\tilde{\pmb v}\rangle{:}_{_{K_0}}=
\langle\tilde{\pmb a},\tilde{\pmb u}\rangle
\langle\tilde{\pmb a},\tilde{\pmb v}\rangle=A(\tilde{\pmb u},\tilde{\pmb v})=
(\tilde{\pmb{u}},\tilde{\pmb{v}})\Big(\frac{1}{2}
\begin{bmatrix}
  0& {}^t\!\tilde{\pmb a}\tilde{\pmb a}\\
  {}^t\!\tilde{\pmb a}\tilde{\pmb a}& 0
\end{bmatrix}\Big)
\begin{bmatrix}
{}^t\!\tilde{\pmb u}\\{}^t\tilde{\pmb v}  
\end{bmatrix}
$$
and the eigenvalue of this rank 2 matrix    
$\begin{bmatrix}
  0& {}^t\!\tilde{\pmb a}\tilde{\pmb a}\\
  {}^t\!\tilde{\pmb a}\tilde{\pmb a}& 0
\end{bmatrix}$
is $\pm\langle\tilde{\pmb{a}},\tilde{\pmb{a}}\rangle$ and $2(m{-}1)$ zeros.  
Hence, $\e_{00}(\tilde{\pmb a})$ has a nontrivial  
$K$-ordered expression for $K$ such  
that $\det\Big(I{-}\begin{bmatrix}
  0& {}^t\!{\tilde{\pmb a}}\tilde{\pmb a}\\
  {}^t\!\tilde{\pmb a}\tilde{\pmb a}& 0
\end{bmatrix}(K{-}K_0)\Big)\not=0$.
In the previous section, we have seen that polar elements behaves 
delicately depending on expression parameters. 
But, we first recall the reason why the double-valued nature of 
$\e_{00}(\tilde{\pmb a})$ appears. 

We explain the reason by using the notaions 
$u=\langle\tilde{\pmb{a}},\tilde{\pmb{u}}\rangle$ and 
$v=\langle\tilde{\pmb{a}},\tilde{\pmb{v}}\rangle$.
There is an adjoint rotation of one parameter subgroups such that 
$$
{\rm{Ad}}(b(s))(e_*^{\frac{t}{i\h}u{\ctt}v}), \quad b(0)=1, \quad 
{\rm{Ad}}(b(\pi))(e_*^{\frac{t}{i\h}u{\ctt}v})=e_*^{-\frac{t}{i\h}u{\ctt}v},
$$
where $1$ at $t=0$ is required by definition of one parameter
subgroups. 
Here, we note that the polar element 
${\e}_{00}=e_*^{\frac{\pi i}{i\h}u{\ctt}v}$ is a member of one
parameter subgroup 
$e_*^{t\frac{1}{\h}u{\ctt}v}$
of crossed symbol. In spite that, if one fixes $t{=}{\pi i}$ first, then  
the normal ordered expression 
${:}{\rm{Ad}}(b(s))(e_*^{\frac{\pi i}{i\h}u{\ctt}v}){:}_{_{K_0}}$ 
is independent of $s$. It follows 
$$
{:}e_*^{\frac{\pi i}{i\h}u{\ctt}v}{:}_{_{K_0}}=
{:}{\rm{Ad}}(b(\pi))(e_*^{\frac{\pi i}{i\h}u{\ctt}v}){:}_{_{K_0}}=
{:}e_*^{-\frac{\pi i}{i\h}u{\ctt}v}{:}_{_{K_0}}.
$$
By the same observation as above, we see that 
${:}\e_{00}(\tilde{\pmb a}){:}_{_{K_0}}
={:}\e_{00}^{-1}(\tilde{\pmb a}){:}_{_{K_0}}$.
On the other hand by the exponential law, we see $\e_{00}(\tilde{\pmb a})$ 
satisfies   
$$
\e_{00}(\tilde{\pmb a})^2=
(\e_{00}^{-1}(\tilde{\pmb a}))^2=-1,\quad 
\e_{00}(\tilde{\pmb a})*\e_{00}^{-1}(\tilde{\pmb a})=1
$$
in the normal ordered expression. 
This was the reason why $\e_{00}(\tilde{\pmb a})$ should be 
regarded as a two valued element.  

In what follows, we show that such double-valued nature is not violated 
by intertwiners.

\subsubsection{Intertwiners are $2$-to-$2$ mappings}


Recalling  \eqref{inter} may be rewritten as 
\begin{equation}
 \label{eq:int003}
I_{_{K}}^{^{K'}}
\Big(\frac{g}{\sqrt{\det(I{-}AK)}};\frac{1}{I{-}AK}A\Big)=
\Big(\frac{g}{\sqrt{\det(I{-}AK')}};\frac{1}{I{-}AK'}A\Big)  
\end{equation}
if $I{-}AK, I{-}AK'$ are invertible. 

Let 
${\mathcal D}_{_K}=
\{A\in{\mathfrak{S}}(2m); \det(I{-}AK)\not=0\}$, and 
let 
${\mathcal D}_{_A}=
\{K\in{\mathfrak{S}}(2m); \det(I{-}AK)\not=0\}$. 

First, we consider the case where $A$ is fixed, then 
$$
\Big\{\Big(\frac{1}{\sqrt{\det(I{-}AK)}};\frac{1}{I{-}AK}A\Big); 
K{\in} {\mathcal D}_{_A}\Big\}
$$
is a double-valued parallel section defined on ${\mathcal D}_{_A}$. 
If $A$ is nonsingular, then
$$
\Big(\frac{\det A}{\sqrt{\det(I{-}AK)}};\frac{1}{I{-}AK}A\Big)=
\Big(\frac{1}{\sqrt{\det(A^{-1}{-}K)}};\frac{1}{A^{-1}{-}K}\Big)
$$
is also a parallel section on on ${\mathcal D}_{_A}$. Taking the 
limit $A^{-1}\to 0$, we have a little strange double-valued parallel section 
\begin{equation}\label{polarpol}
\Big(\frac{1}{\sqrt{\det({-}K)}};-\frac{1}{K}\Big), \quad 
K\in {\mathcal D}_{\infty}, 
\end{equation}
where ${\mathcal D}_{\infty}=\{K; \det K\not=0\}$. If $K=K_0$, then
this is 
$\frac{1}{\sqrt{(-1)^m}}e^{-\frac{2}{i\h}\sum_k u_kv_k}.$
Hence, \eqref{polarpol} may be regarded as the $K$-ordered expression 
of the total polar element \eqref{polarpolar00}.

\bigskip
Next, we consider the case where $K$ is fixed and $A$ is moving, 
then the space 
$$
{\widetilde{\mathcal D}}_{_K}{=}
\Big\{\Big(\frac{1}{\sqrt{\det(I{-}AK)}};\frac{1}{I{-}AK}A\Big); 
A{\in} {\mathcal D}_{_K}\Big\}
$$  
for $K\not=0$ is viewed as a nontrivial double cover of the space 
${\mathcal D}_{_K}$. ${\widetilde{\mathcal D}}_{0}$ for $K=0$ is viewed as 
$\{(\pm 1; A); A{\in} {\mathcal D}_0\}$. 
Let $\pi_{_K}$ be the natural projection, and let ${\mathcal D}_{_{KK'}}{=}
{\mathcal D}_{{_K}}{\cap}\,{\mathcal D}_{_{K'}}$.
\begin{prop}
 \label{paramean}
The intertwiner $I_{_{K}}^{^{K'}}$ is then a 
$2$-to-$2$ mapping from ${\widetilde{\mathcal D}}_{_K}$ to 
${\widetilde{\mathcal D}}_{_{K'}}$. Hence, the intertwiner keeps the
double-valued nature of the ${*}$-exponential functions of quadratic forms.
\end{prop}

Precisely speaking, the intertwiner $I_{_{K}}^{^{K'}}$ is defined as a 
mapping of $\pi_{_K}^{-1}{\mathcal D}_{_{KK'}}$ onto 
$\pi_{_{K'}}^{-1}{\mathcal D}_{_{K'K}}$. Since 
the transformation 
$T_{_K}^{^{K'}}: \frac{1}{1{-}AK}A \to \frac{1}{1{-}AK'}A$ changes the 
homotopical nature of closed curves via the movement of singularities,   
the notion of ``lift'' of closed curves by parallel displacement along 
 closed curves is not stable. Recall again that these arguments have 
nothing  to do with the Weyl algebra. 

\noindent
{\bf Note}\,\,\,Intertwiners fails the cocycle condition as one-to-one 
mappings, i.e. 
$I_{_{K''}}^{^{K}}I_{_{K'}}^{^{K''}}I_{_{K}}^{^{K'}}$ 
may not equal $1$, but it is ${\pm}1$ for exponential 
functions of quadratic forms. This is similar to 
${\mathbb Z}_2$-gerbes.

If $g$ is fixed｡､${\widetilde{\mathcal D}}_{_K}$ is a double 
covering space of ${\mathcal D}_{_K}$｡･As in the case of one 
variable, take the following diagram in mind｡ｧ
$$
\begin{matrix}
\widetilde{\mathcal D}_{_K}& \supset & 
\pi^{-1}({\mathcal D}_{_K}{\cap}{\mathcal D}_{_{K'}})&
\overset{I_{_K}^{^{K'}}}{\longrightarrow}&
\pi^{-1}({\mathcal D}_{_{K'}}{\cap}{\mathcal D}_{_K})&
\subset &\widetilde{\mathcal D}_{_{K'}}\\
\downarrow\,\pi&  &\downarrow\,\pi&  &\downarrow\,\pi&  
&\downarrow\,\pi \\
{\mathcal D}_{_K}& \supset & 
{\mathcal D}_{_K}{\cap}{\mathcal D}_{_{K'}}&
=\!=&
{\mathcal D}_{_{K'}}{\cap}{\mathcal D}_{_K}&
\subset&{\mathcal D}_{_{K'}}.
\end{matrix}   
$$
Turning around the circle in the picture of the 
l.h.s., the sign does not change as 
$\circ$ is not a singular point. However, turning around the circle 
in the r.h.s., the sign changes as $\bullet$ is a branched 
singular point.  Similarly, as the $\bullet$ in the l.h.s.
picture is a branched singular point, the sign changes around 
this point. Hence, provided $c$ is a constant, 
$\pi^{-1}(p)$ must be two points, which one cannot distinguish,   
for these two points exchange each other when one goes around 
the point $\bullet$. On the other hand, since $\circ$ is not a 
singular point in the r.h.s., these two point can be 
distinguished. 

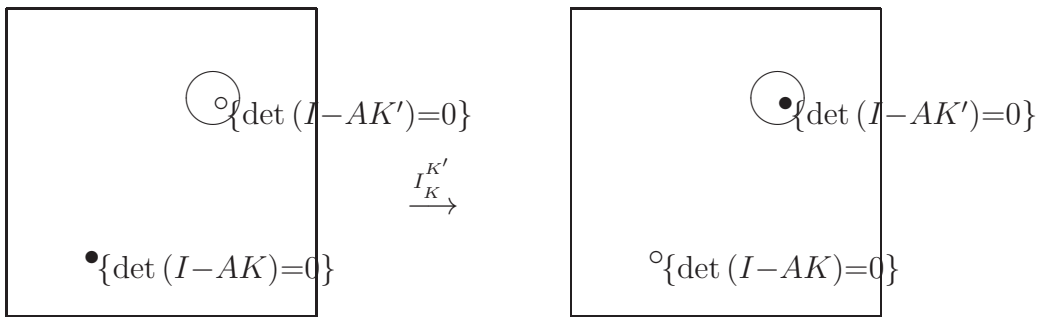
\setlength{\unitlength}{.34mm}
\begin{picture}(140,120)(0,-20)
\thinlines
\put(0,0){\line(1,0){120}}
\put(0,0){\line(0,1){120}}
\put(120,0){\line(0,1){120}}
\put(0,120){\line(1,0){120}}
\put(30,20){$\bullet$}
\put(35,15){$\{\det{(I{-}AK)}{=}0\}$}
\put(80,80){$\circ$}
\put(85,75){$\{\det{(I{-}A{K'})}{=}0\}$}
\put(80,85){\circle{20}}
\put(155,40){$\overset{I_{_K}^{^{K'}}}{\longrightarrow}$}
\end{picture}
\qquad
\begin{picture}(180,160)(-50,-20)
\thinlines
\put(0,0){\line(1,0){120}}
\put(0,0){\line(0,1){120}}
\put(120,0){\line(0,1){120}}
\put(0,120){\line(1,0){120}}
\put(30,20){$\circ$}
\put(35,15){$\{\det{(I{-}AK)}{=}0\}$}
\put(80,80){$\bullet$}
\put(85,75){$\{\det{(I{-}A{K'})}{=}0\}$}
\put(80,85){\circle{20}}
\end{picture}

Consequently, one cannot trace how points of 
$\pi^{-1}({\mathcal D}_{_K}{\cap}{\mathcal D}_{_{K'}})$
map onto points of 
$\pi^{-1}({\mathcal D}_{_{K'}}{\cap}{\mathcal D}_{_{K}})$ 
with one-to-one correspondence. In spite of this 
difficulty, one can trace this mapping as a 2-to-2 
mapping. If one views these two points 
as a singleton, then  this produces nothing but the 
identification of $\widetilde{\mathcal D}_{_K}$ 
with ${\mathcal D}_{_K}$, and the mapping is nothing but 
the identity mapping of ${\mathcal D}_{_K}$ onto 
${\mathcal D}_{_{K'}}$. 

Since this procedure loses much information, we prefer to regard 
such a mapping as a $2$-to-$2$ mapping, for these two points 
can be distinguished locally.  

Moreover, one can define a kind of group operation 
via the definition of $2$-to-$2$ mappings.
Even in such a situation, some partial area of object 
one may fix a unit, inverse and product in a univalent 
way to obtain a genuine group. Hence, local differential 
geometry can be done without any difficulty.

To treat elements with double-valued nature, we have to discuss the intertwiners
to generic ordered expressions.   
Setting ${\pmb u}{=}(\tilde u_1,\dots,\tilde u_m, \tilde v_1,\dots,\tilde v_m)$ and recalling 
$K_0{=}
\begin{bmatrix}
0&I_m\\
I_m& 0\\ 
\end{bmatrix}$, the formula \eqref{eq:norexpm2} 
is rewritten as 
\begin{equation}\label{expmatrix}
{:}{\rm{exp}}_*{\frac{s}{2i\h}
\langle{\pmb u}
\begin{bmatrix}
0&C\\
{}^tC&0\\
\end{bmatrix},
{\pmb u}\rangle}
{:}_{_{K_0}}{=}
e^{\frac{s}{2}{\rm{Tr}}C}
{\rm{exp}}{\frac{1}{2i\h}
\langle{\pmb u}
\begin{bmatrix}
0&e^{sC}{-}I\\
e^{s{}^t\!C}{-}I&0\\
\end{bmatrix},
{\pmb u}\rangle}.
\end{equation}
In particular, we see 
$$
{:}e_*^{\frac{s}{i\h}
(\tilde u_1{\ctt}\tilde v_1{+}{\cdots}{+}\tilde u_m{\ctt}\tilde v_m)}{:}_{_{K_0}}{=}
e^{\frac{ms}{2}}
{\rm{exp}}{\frac{1}{2i\h}
\langle{\pmb u}
\left[\small\begin{matrix}
0&e^{sI}{-}I\\
e^{sI}{-}I&0\\
\end{matrix}\right],
{\pmb u}\rangle}.
$$
$$
{:}e_*^{\frac{2\pi i}{i\h}
(\tilde u_1{\ctt}\tilde v_1{+}{\cdots}{+}\tilde u_m{\ctt}\tilde v_m)}{:}_{_{K_0}}{=}
(-1)^m, \quad 
{:}e_*^{\frac{\pi i}{i\h}
(\tilde u_1{\ctt}\tilde v_1{+}{\cdots}{+}\tilde u_m{\ctt}\tilde v_m)}{:}_{_{K_0}}{=}
i^m{\rm{exp}}{\frac{{-}1}{i\h}
\langle{\pmb u}
\left[\begin{matrix}
0&I\\
I&0\\
\end{matrix}\right],
{\pmb u}\rangle}.
$$
The sign ambiguity of $\sqrt{\,\,}$ does not appear on this 
expression, but precisely speaking we should write the l.h.s. 
$$
{:}e_*^{[0\to 2\pi i]\frac{1}{i\h}
(\tilde u_1{\ctt}\tilde v_1{+}{\cdots}{+}\tilde u_m{\ctt}\tilde v_m)}{:}_{_{K_0}},  \quad 
{:}e_*^{[0\to\pi i]\frac{1}{i\h}
(\tilde u_1{\ctt}\tilde v_1{+}{\cdots}{+}\tilde u_m{\ctt}\tilde v_m)}{:}_{_{K_0}},
$$
where $[0{\to}a]$ implies the path given by the straight line segment. 

\medskip 
It is very natural to expect that there is $K$ such that 
$$
\hat\e_{00}(\tilde{\pmb a}){*_{_K}}\hat\e_{00}(\tilde{\pmb b})=
-\hat\e_{00}(\tilde{\pmb b}){*_{_K}}\hat\e_{00}(\tilde{\pmb a}),\qquad 
(\text{if }\,\,\,\langle\tilde{\pmb a},\tilde{\pmb b}\rangle=0).
$$

\subsection{Clifford algebras in 
$(H\!ol({\mathbb C}^{2m}),  {*}_{_K})$}\label{CiffCliff}

Since the group $Spin(m)$ is usually constructed as a 
Clifford algebra, it is natural to think that the group ring 
of the group ${\widetilde{SO}}(m)$ under some other expression
parameter $K$ considered in $(H\!ol({\mathbb C}^n),{*_{K}})$  
has the structure of Clifford algebra. 

At this moment, this is supported only 
by the following strange phrase:
\begin{center}
\fbox{\parbox[c]{.70\linewidth}{Since the ${\e}_{00}(k)$'s are defined as double-valued elements, 
the identities  
$$
{\e}_{00}(k)=-{\e}_{00}(k),\quad 
{\e}_{00}(k){*}{\e}_{00}(\ell)
{=}-{\e}_{00}(\ell){*}{\e}_{00}(k)
$$
do not contradictory}. 
}
\end{center} 

The goal of this section is the following. 
\begin{thm}\label{EXPspecial} 
There is an expression parameter $K_s$ having the following properties: 
Let $V_*$ be $1$ or any partial polar element without involving 
${\e}_{00}(k), \,\,{\e}_{00}(\ell)$ $(k,\ell)$ such that $k\not=\ell$.
Then,  
$$
\begin{aligned}
&{:}{\e}_{00}(k)^2{*}V_*{:}_{_{K_s}}{=}-{:}V_*{:}_{_{K_s}}\\
&{:}({\e}_{00}(k){*}{\e}_{00}(\ell))^2_*{*}V_*{:}_{_{K_s}}=-{:}V_*{:}_{_{K_s}}.
\end{aligned}
$$
\end{thm}  

Since the identity above gives
$$
{:}{\e}_{00}(k){*}{\e}_{00}(\ell){:}_{_{K_s}}=
-{:}{\e}_{00}(\ell)^{-1}{*}{\e}_{00}(k)^{-1}{:}_{_{K_s}},
$$
by noting that ${\e}_{00}(k)^{-1}=\pm {\e}_{00}(k)$ by the double-valued
nature, but the $\pm$-sign can be controlled to be independent of $k$ 
(cf \eqref{matrices}), we see that  
$$
{:}{\e}_{00}(k){*}{\e}_{00}(\ell){:}_{_{K_s}}=
-{:}{\e}_{00}(\ell){*}{\e}_{00}(k){:}_{_{K_s}}.
$$

Under such an ordered expression $K=K_s$, the system  
$$
p(u,v){*_{_K}}{\e}_{00}(1)^{\e_1}{*_{_K}}{\cdots}{*_{_K}}{\e}_{00}(m)^{\e_m}
$$
naturally forms an algebra under the $*_{_{K_s}}$-product,
which may be
called the {\bf Weyl-Clifford algebra}.  This means the 
super-theoretic expressions are already built in the extended Weyl
algebra. That is, we have no need to construct a {\it new}
mathematical theory to absorb the super manifold theory. 

\medskip
The next three steps are essential for the proof of Theorem\,\ref{EXPspecial}: 

\noindent
(1:) Note first that in the normal ordered expression, 
partial polar elements form a commutative 

algebra. Moreover, we already see that    
$$
{:}e_*^{\frac{it_1}{i\h}u_1{\ctt}v_1}{*}\cdots{*}
 e_*^{\frac{it_m}{i\h}u_m{\ctt}v_m}{:}_{_{K_0}}=
{:}e_*^{\frac{1}{i\h}(it_1u_1{\ctt}v_1{+}\cdots{+}it_m u_m{\ctt}v_m)}{:}_{_{K_0}}.
$$

\noindent
(2:) Let $V_*$ be any partial polar element or $\pm 1$.
Applying the intertwiner $I_{_{K_0}}^{^{K_s}}$ to this system, 

we show the following:

\noindent 
(2:1) $I_{_{K_0}}^{^{K_s}}(e_*^{\frac{it}{i\h}u_k{\ctt}v_k}{*_{_{K_0}}}V_*)$ has
no singular point on the interval $[0,\pi]$. 

\noindent 
(2:2) In spite of this, if $V_*$ does not contain 
$e_*^{\frac{it}{i\h}u_k{\ctt}v_k}$, 
$e_*^{\frac{it}{i\h}u_{\ell}{\ctt}v_{\ell}}$, then 
$I_{_{K_0}}^{^{K_s}}
(e_*^{\frac{it}{i\h}(u_k{\ctt}v_k{+}u_{\ell}{\ctt}v_{\ell}}{*_{_{K_0}}}V_*)$
has singular 

points on the open intervals $(0,\pi)$ and $(\pi,2\pi)$. 

%

\noindent 
(3:) If $V_*$ does not contain $e_*^{\frac{it}{i\h}u_k{\ctt}v_k}$, 
$e_*^{\frac{it}{i\h}u_{\ell}{\ctt}v_{\ell}}$, 
then 
${:}e_{*}^{\frac{i}{i\h}
(su_k{\ctt}v_k{+}tu_{\ell}{\ctt}v_{\ell})}{*}V_*{:}_{_K}$ has no
singular point in 

$[0,\pi]{\times}[0,\pi]$ except on the diagonal set $s{=}t$.
 
\bigskip

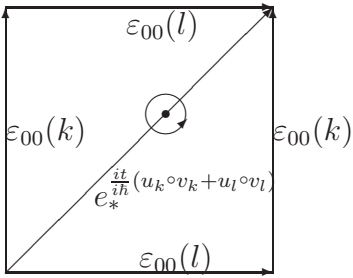
\mbox{
\setlength{\unitlength}{10pt}
\begin{picture}(13, 10)(2,-1)
\put(0,0){\vector(1,1){10}}
\put(0,0){\vector(1,0){10}}
\put(0,0){\vector(0,1){10}}
\put(10,0){\vector(0,1){10}}
\put(0,10){\vector(1,0){10}}
\put(6,6){\circle{1.5}}
\put(6.7,5.7){\vector(1,1){.1}}
\put(6,6){\circle*{0.3}}
\put(0,5){${\e}_{00}(k)$}
\put(10,5){${\e}_{00}(k)$}
\put(3.3,2.5){$e_*^{\frac{it}{i\h}
(u_k{\ctt}v_k{+}u_l{\ctt}v_l)}$}
\put(5,.2){${\e}_{00}(l)$}
\put(4.5,9){${\e}_{00}(l)$}
\end{picture}}
\hfill
\parbox[b]{.65\linewidth}
{The proof of Theorem\,\ref{EXPspecial} is given as follows: 
Suppose $V_*$ does not contain ${\e}_{00}(k)$,\,${\e}_{00}(\ell)$.
Note that ${:}e_{*}^{\frac{i}{i\h}
(su_k{\ctt}v_k{+}tu_{\ell}{\ctt}v_{\ell})}{*}V_*{:}_{_K}$
has no singular point on the lower triangular domain 
$\{(s,t); 0\leq t<s\leq\pi\}$. There is one singular 
point at $(\mu,\mu)$. Therefore, we see    
${:}({\e}_{00}(k){*}{\e}_{00}(l)){*}V_*{:}_{_K}$ equals  
${:}e_{*}^{\frac{it}{i\h}
(u_k{\ctt}v_k{+}u_{\ell}{\ctt}v_{\ell})}{*}V_*{:}_{_K}$ 
defined by taking the path avoiding the singular point 
anti-clockwise. 
Similarly, ${:}({\e}_{00}(l){*}{\e}_{00}(k)){*}V_*{:}_{_K}$
 equals to the element ${:}e_{*}^{\frac{it}{i\h}
(u_k{\ctt}v_k{+}u_{\ell}{\ctt}v_{\ell})}{*}V_*{:}_{_K}$ 
which is defined by taking the clockwise path 
avoiding the singular point.}  

Note here that products such as  
${:}e_{*}^{\frac{si}{i\h}u_k{\ctt}v_k}{*}V_*{:}_{_K}$,  
${:}e_{*}^{\frac{si}{i\h}
(u_k{\ctt}v_k{+}u_{\ell}{\ctt}v_{\ell})}{*}V_*{:}_{_K}$ are defined
by solving the evolution equation with initial data $V_*$. Such a
procedure for constructing products will 
be called {\it the path  connecting product}.

\noindent
{\bf Singularity makes change of sign.}\,\, Here, we show that 
the singularity makes the change of sign, and hence 
\begin{equation}\label{exchange}
{:}{\e}_{00}(k){*}{\e}_{00}(l){*}V_*{:}_{_K}
{=}{-}{:}{\e}_{00}(l){*}{\e}_{00}(k){*}V_*{:}_{_K},
\end{equation} 
where it is assumed that  $V_*$ does not contain ${\e}_{00}(k)$,\,${\e}_{00}(\ell)$. 

\bigskip
Consider the product $e_{*}^{sH_*}{*}e_*^{tK_*}$ for two 
quadratic forms $H_*, K_*$ in $(s,t)\in {\mathbb C}^2$ such that $[H_*,K_*]=0$. 
In our situation it may be assumed  
$e_{*}^{sH_*}{*}e_*^{tK_*}=\pm e_*^{tK_*}{*}e_{*}^{sH_*}$ 
with the sign ambiguity, that is, the phase parts of both sides coincides and 
the sign ambiguity appears only in the amplitude parts. 

In general, 
$e_{*}^{sH_*}{*}e_*^{tK_*}$ has a singular set $S$ of complex 
codimension 1. We see that the origin $(0,0)$ is not contained in $S$. 
Since $S$ is a branched singularity, we have to prepare two 
sheets ${\mathbb C}_+^2$, ${\mathbb C}_-^2$ and ``slit'' $\Sigma$ 
of real codimension 1 to connect these two sheets. $\Sigma$ is set so that 
${\mathbb C}^2{\setminus}\Sigma$ is simply connected 
and there is no singular point. 

Now, restrict the parameter $(s,t)\in {\mathbb R}^2$
in $e_{*}^{sH_*}{*}e_*^{tK_*}$. 
One may assume that ${\mathbb R}^2$ is transversal to $S$ 
in generic ordered expression. Hence, if 
$S\cap{\mathbb R}^2\not=\emptyset$, then this is a 
discrete set and  $\Sigma\cap{\mathbb R}^2$ is a collection of 
(real one dimensional) curves starting at a singular point 
ending another singular point or $\infty$.

\noindent
\setlength{\unitlength}{.3mm}
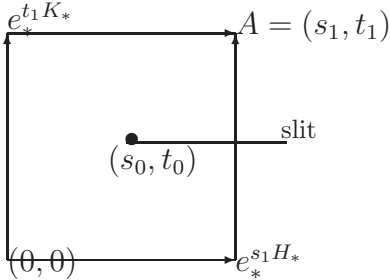
\begin{picture}(150,130)(-10,-15)
\thinlines
\put(0,0){\vector(1,0){100}}
\put(0,0){\vector(0,1){100}}
\put(100,0){\vector(0,1){100}}
\put(0,100){\vector(1,0){100}}
\put(0,-5){$(0,0)$}
\put(100,-5){$e_{*}^{s_1H_*}$}
\put(0,102){$e_{*}^{t_1K_*}$}
\put(100,100){$A=(s_1,t_1)$}
\put(51,50){$\bullet$}
\thicklines
\put(52,52){\line(1,0){70}}
\put(120,54){\footnotesize{slit}}
\put(44,40){$(s_0,t_0)$}
\end{picture}
\hfill
\parbox[b]{.5\linewidth}
{As we have two sheets, there are two ``origin'', 
$(0,0){\in}{\mathbb C}_+^2$ and $(0,0){\in}{\mathbb C}_-^2$. 
Since $e_{*}^{0H_*}{*}e_*^{0K_*}$ is $1$ in the positive sheet 
${\mathbb C}_+^2$, the origin in the negative sheet must 
be treated as $-1$.  Now, consider $e_{*}^{s_1H_*}{*}e_*^{t_1K_*}$ 
and  $e_*^{t_1K_*}{*}e_{*}^{s_1H_*}$. The first one is defined by 
by the solution of the evolution equation 
$$
\frac{d}{dt}f_t=H_*{*}f_t,\quad f_0=e_*^{t_1K_*}. 
$$
}

\noindent 
We indicate this by the notation 
$e_{*}^{[0\to s_1]H_*}{*}e_*^{t_1K_*}$. This is the clockwise chasing
from the origin. On the contrary,  
$e_*^{[0\to t_1]K_*}{*}e_{*}^{s_1H_*}$ means the anti-clockwise 
chasing from the origin. Now suppose there is a singular point 
$(s_0,t_0)$ and a slit as it is seen in the left figure, then 
$e_*^{[0\to t_1]K_*}{*}e_{*}^{s_1H_*}$ is lying in the opposite 
sheet. By this way, the sign changes around 
a singular point.

\subsubsection{Intertwiners to generic ordered expressions}

In the first part of discussions, we recall  \eqref{inter}.
Via the intertwiner $I_{_{K_0}}^{^K}$  from \eqref{expmatrix}, 
the $K$-expression of the $*$-exponential function 
$e_{*}^{\frac{t}{i\hbar}\sum C_{kl}u_k{\ctt}v_l}$
is easily obtained.   
Recall ${:}e_*^{\frac{i}{i\h}
(t_1u_1{\ctt}v_1{+}\cdots{+}t_mu_m{\ctt}v_m)}{:}_{_{K_0}}$ 
is an entire function of $(t_1,\cdots,t_m)$, $t_i{\in}{\mathbb C}$, 
in the normal ordered expression which is 
written as 
$
e^{\frac{i}{2}(t_1{+}\cdots{+}t_m)}
e^{\frac{1}{i\h}\langle{\pmb u}A,{\pmb u}\rangle}$, where 
$$
A{=}
\begin{bmatrix}
0&C\\
C&0
\end{bmatrix},\quad 
C{=}{diag}(\tau_1,\cdots,\tau_m), \quad 
\tau_k{=}\frac{1}{2}(e^{it_k}{-}1), \quad {\rm{cf}}. \eqref{eq:norexpm2}. 
$$

\medskip
In this section, we use special ordered expressions $K_{s}$,  
where $K_{s}$ is given step by step 
by \eqref{specialorder00}, \eqref{ordering}, 
 \eqref{matrices} and \eqref{keycond} as follows: 
We set 
\begin{equation}\label{specialorder00}
K_s{=}
\begin{bmatrix}
S&T\\
T&S
\end{bmatrix},  
\quad
{}^tS{=}S
\quad
{}^tT{=}T,\quad S,\, T {\in} M(m,{\mathbb C}).
\end{equation} 

By \eqref{inter}, we need to know $\sqrt{\det(I{-}A(K{-}K_0))}$. 
$\det(I{-}A(K{-}K_0))$ is given by elementary transformation as follows:  
\begin{equation}\label{elemtrf}
\begin{vmatrix}
I{-}C(T{-}I)&{-}CS\\
{-}CS& I{-}C(T{-}I)
\end{vmatrix}{=}
\det\big(I{-}C(T{+}S{-}I)\big)
\det\big(I{-}C(T{-}S{-}I)\big).
\end{equation}
Note that $S{+}T{=}U, S{-}T{=}V$ are arbitrary symmetric matrices.

\medskip
As we want to use a $K$-ordered expression such that 
$$
{\rm{sgn}}({:}e_*^{\frac{t}{i\h}
\tilde u_i{\ctt}\tilde v_i}{:}_{_{K}}){=}
{\rm{sgn}}({:}e_*^{\frac{t}{i\h}
\tilde u_j{\ctt}\tilde v_j}{:}_{_{K}})
$$
for every $i,j$, we restrict $K$ to symmetric matrices such that 
\begin{equation}\label{ordering}
K_{s}{=}
\begin{bmatrix}
i\rho I&cI\\
cI&i\rho I
\end{bmatrix}{+}
\begin{bmatrix}
S'&T'\\
T'&S'
\end{bmatrix},\quad \rho, c\in {\mathbb R}, 
\end{equation}
where the diagonal components of $S', T'$ are zero, and 
all other entries are the same complex constant.  
It is easy to see that the formula for 
${:}f_*(\tilde u_j,\tilde v_j){:}_{_{K}}$ is written by replacing 
$i$ by $j$ in the formula for ${:}f_*(\tilde u_i,\tilde v_i){:}_{_{K}}$, 
since we use only $i\rho$ and $c$ in the computation of 
${:}f_*(\tilde u_i,\tilde v_i){:}_{_{K}}$. 

In what follows we set that 
\begin{equation}\label{matrices}
T'{=}
\begin{bmatrix}
0&a&a&\cdots&a\\
a&0&a&\cdots&a\\
a&a&0&\cdots&a\\
\vdots&\vdots&\ddots&\ddots&\vdots\\
a&a&a&\cdots&0\\
\end{bmatrix}, \quad 
S'{=}
\begin{bmatrix}
0&ib&ib&\cdots&ib\\
ib&0&ib&\cdots&ib\\
ib&ib&0&\cdots&ib\\
\vdots&\vdots&\ddots&\ddots&\vdots\\
ib&ib&ib&\cdots&0\\
\end{bmatrix}, \quad a, b \in \mathbb R, 
\end{equation}
so that $T{-}S=\overline{T{+}S}$. 
Further, in \eqref{keycond} below  we put the additional condition
that $c=a>0$ and $\rho>b$. Hence, $T$ in \eqref{specialorder00} is a
matrix such that $T_{ij}=c>0$.

We refer to $K_{s}$ as the {\bf special ordered expression}, or $K_{s}$-expression.

\subsubsection{Vertexes and 2-dimensional nets}\label{2nets}  

To establish the product formula for polar elements in 
general $K$-ordered expressions, we have to prepare 
several tools used in the definition of  
products. 

\bigskip
Denote by $(t_1,t_2,\cdots,t_m)$ a point of ${\mathbb R}^m$. 
The {\it lattice point} is the subset $(\pi{\mathbb Z})^m$  of
${\mathbb R}^m$, and the {\it 1-dimensional lattice} is the subset of ${\mathbb R}^m$
such that only one of $(t_1,t_2,\cdots,t_m)$ is in $\mathbb R$ and 
others are $k\pi$, $k{\in}{\mathbb Z}$.   
A {\it vertex} is a point $(\delta_1,\cdots,\delta_m)$ 
where $\delta_i{=}$ $0$ or $\pi$.   
The number of $\pi's$ is called the {\it index of the vertex}.

The {\it 2-dimensional lattice} is the subset of ${\mathbb R}^m$ 
with only two of $(t_1,t_2,\cdots,t_m)$ in $\mathbb R$ and all
others $k\pi$, $k{\in}{\mathbb Z}$. Denote the 
2-dimensional lattice, the 1-dimensional lattice and the set of lattice points 
by $L_m(2)$, $L_m(1)$, $L_m(0)$ respectively. 

We often use $(t_1,t_2,\cdots,t_m)\in{\mathbb R}^m$ to indicate the
$*$-exponential function 
$$
{:}e_*^{\frac{i}{i\h}
(t_1u_1{\ctt}v_1{+}\cdots{+}t_{m}u_m{\ctt}v_m)}{:}_{_{K}}.
$$
Hence, {\it lattice points} in $K_0$-ordered expression are 
$$
{:}L_*{:}_{_{K_0}}{=}
{:}e_*^{\frac{i}{i\h}(\delta_1\tilde u_{i_1}\ctt \tilde v_{i_1}{+}\cdots{+}
\delta_m \tilde u_{m}\ctt \tilde v_{m})}{:}_{_{K_0}},\quad 
\delta_i{=}0\,\,{\text{or}}\,\,\pi\ell.
$$ 

The next proposition is a basic result proved by the uniqueness of the real
analytic solutions of evolution equations.
\begin{prop}\label{fundament} 
For a lattice point $L_*$, if 
${:}e_{*}^{\frac{it}{i\h}\tilde u_k{\ctt}\tilde v_k}{*}L_{*}{:}_{_{K}}$ 
is not singular on $t{\in}[0,\pi]$, then 
${:}{\e}_{00}(k){*}L_{*}{:}_{_{K}}$ is defined as a 
single element. This gives the $K$-ordered expression of 
vertex ${\e}_{00}(k){*}L_{*}$.  
\end{prop}

We often use the variable $\tau=\frac{1}{2}(e^{it}{-}1)$ or
$\tau^{-1}$ instead of $t$, when the variables $e^{it}$ is restricted
in the unit circle.  

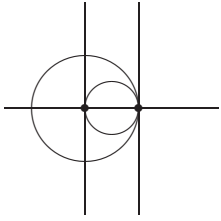
\mbox{
\setlength{\unitlength}{1pt}
\begin{picture}(80,80)(-20,-40)
\put(20,0){\circle{40}}
\put(20,0){\circle*{3}}
\put(40,0){\circle*{3}}
\put(30,0){\circle{20}}
\put(40,-40){\line(0,1){80}}
\put(20,-40){\line(0,1){80}}
\put(-10,0){\line(1,0){80}}
\end{picture}}
\hfill
{\parbox[b]{.6\linewidth}
{Note that $\tau_i^{-1}{+}1$ is a negative (resp. positive) 
pure imaginary number if   
$t_i{\in}(0,\pi)$ (resp. $t_i{\in}(-\pi,0)$).
To see this quickly, let $D_1$ be the 
unit disk at the origin. Then, $\frac{1}{2}(1{+}D_1)$ is the 
disk of radius $\frac{1}{2}$ with the center at $\frac{1}{2}$.
Hence, its inverse is the right half-plane with ${\rm{Re}}z{>}1$.
We use this argument very often in the calculation. Note also 
that when $\tau_i^{-1}{+}1$ are used in the calculation, 
$t_i{=}2\pi i$ corresponds to $\tau_i^{-1}{+}1{=}\infty$.}

Note that 
$$
\tau=\frac{e^{i0}{-}1}{2}{=}0,\,\,{\text{hence}}\,\,
\tau^{-1}{+}1{=}\infty, \quad 
\text{and}\quad
\tau=\frac{e^{\pm i\pi}{-}1}{2}{=}{-}1,\,\,{\text{hence}}\,\,
\tau^{-1}{+}1{=}0. 
$$

\medskip
To obtain a single-valued product formula between partial polar 
elements, we have to consider elements 
${:}e_*^{\frac{i}{i\h}
(t_1\tilde u_1{\ctt}\tilde v_1{+}\cdots{+}t_m\tilde u_m{\ctt}\tilde v_m)}{:}_{_{K}}$
where some of 
$(t_1,\cdots, t_m)$ are $0$ or $\pm\pi$. 
Say $t_{i_1},\cdots, t_{i_k}$ 
are $0$, and $t_{j_1},\cdots, t_{j_{\ell}}$ are $\pm\pi$. 
By a suitable change of rows and columns, we can assume 
$\tau_1=\cdots{=}\tau_{k}{=}0$ without loss of generality, and 
the computation of the determinant is reduced to the case 
$(m{-}k){\times}(m{-}k)$-matrices, where 
$C{=}diag(\tau_1,\cdots,\tau_{m{-}k{-}\ell}, 
\pm\pi,\cdots,\pm\pi)$
 with $\sharp\,(\pm\pi)=\ell$ (the index of the vertex).  

Since $\tau_i{\not=}0$ in the reduced matrices, we have, 
by setting $\sigma{=}m{-}k{-}\ell$,  
$$
\begin{aligned}
\det\big(I{-}&C(T{+}S{-}I)\big)
{\det\big(I{-}C(T{-}S{-}I)\big)}\\
&{=}(\det C)^2
\det\big(C^{-1}{+}I{-}(T{+}S)\big)
{\det\big(C^{-1}{+}I{-}(T{-}S)\big)}\\
&{=}
(\det C)^2\det\Big(((diag(\tau^{-1}_1{+}1,\cdots,
\tau^{-1}_{\sigma}{+}1,0,\cdots,0)){-}(T{+}S))\Big)\\
&\qquad\qquad
{\times}\det\Big(((diag(\tau^{-1}_1{+}1,\cdots,
\tau^{-1}_{\sigma}{+}1,0,\cdots,0)){-}(T{-}S))\Big),
\end{aligned}
$$
where the number of $0's$ is  $\sharp 0 = \ell$. 

\medskip

For simplicity, we set $\alpha{=}c{+}i\rho$, $\beta{=}a{+}ib$. 
Hence, the determinant is decomposed into two 
factors $F{\times}{\overline{F}}$, and one of them is    
given by the formula as follows:  
$$
\begin{vmatrix}
\tau^{-1}{+}1{+}\alpha&\beta&\cdots&\beta\\
\beta&\alpha&\cdots&\beta\\
\vdots&\vdots&\ddots&\vdots\\
\beta&\beta&\cdots&\alpha
\end{vmatrix}
{=}
\begin{vmatrix}
\tau^{-1}{+}1&0&\cdots&0\\
\beta&\alpha&\cdots&\beta\\
\vdots&\vdots&\ddots&\vdots\\
\beta&\beta&\cdots&\alpha
\end{vmatrix}
{+}
\begin{vmatrix}
\alpha&\beta&\cdots&\beta\\
\beta&\alpha&\cdots&\beta\\
\vdots&\vdots&\ddots&\vdots\\
\beta&\beta&\cdots&\alpha
\end{vmatrix}
$$
$$
{=}
(\tau^{-1}{+}1)
\begin{vmatrix}
1&0&\cdots&0\\
0&\alpha&\cdots&\beta\\
\vdots&\vdots&\ddots&\vdots\\
0&\beta&\cdots&\alpha
\end{vmatrix}
{+}
\begin{vmatrix}
\alpha&\beta&\cdots&\beta\\
\beta&\alpha&\cdots&\beta\\
\vdots&\vdots&\ddots&\vdots\\
\beta&\beta&\cdots&\alpha
\end{vmatrix}
$$
For $\ell{=}0$, this is given by $\tau^{-1}{+}1{+}\alpha$, and for 
$\ell{\geq}1$, it is     
$$
\begin{aligned}
(\tau^{-1}{+}1)&(\alpha{+}(\ell{-}1)\beta)(\alpha{-}\beta)^{\ell{-}1}
{+}
(\alpha{+}{\ell}\beta)(\alpha{-}\beta)^{\ell}\\
&=\Big((\tau^{-1}{+}1)(\alpha{+}(\ell{-}1)\beta){+}
(\alpha{+}{\ell}\beta)(\alpha{-}\beta)\Big)
(\alpha{-}\beta)^{\ell{-}1}.
\end{aligned}
$$ 

We now assume the following 
\begin{equation}\label{keycond}
{\rm{Re}}(c{+}i\rho{-}(a{+}ib)){=}0, \quad 
{\rm{Im}}(c{+}i\rho{-}(a{+}ib)){>}0, \quad c>0.
\end{equation}
In particular, this implies $c{=}a{>}0$. Note that 
we have three dimensions of freedom for
$(a, b, \rho)$. 

\bigskip
Since $(\alpha{-}\beta){\not=}0$ by \eqref{keycond}, 
the determinant vanishes if and only if  
$$
-(\tau^{-1}{+}1){=}
(\alpha{-}\beta)(1{+}\frac{\beta}{\alpha{+}(\ell{-}1)\beta}),
\quad 
{=}\alpha,\,(\ell{=}0), \quad {\text{or}}\,\,
 \in i{\mathbb R}_+.
$$

\begin{lem}\label{estimate1} 
Under the assumption \eqref{keycond}, we see that  
${\rm{Re}}(\alpha){>}0$ for $\ell{=}0$, and for $\ell{\geq}1$,
$${\rm{Re}}
\Big((\alpha{-}\beta)
(1{+}\frac{\beta}{\alpha{+}(\ell{-}1)\beta})\Big){>}0. 
$$
Thus, there is no singular point on the 
domain ${\rm{Re}}(\tau^{-1}{+}1)\geq 0$. 

${:}e_*^{\frac{it}{i\h}u{\ctt}v}{:}_{_K}$ is a 
complex semigroup on the upper half-plane, and alternating 
$2\pi$-periodic and reflection symmetric. There are
no singular points on the imaginary axis.
\end{lem}

\noindent
{\bf Proof}. The case $\ell{=}0$ is trivial. 
Since $\alpha{-}\beta$ is positive pure imaginary
 under \eqref{keycond}, we easily see  
${\rm{Re}}
\Big((\alpha{-}\beta)
(1{+}\frac{\beta}{\alpha{+}(\ell{-}1)\beta})\Big){<}0$ 
is equivalent to  
${\rm{Im}}(\frac{\beta}{\alpha{+}(\ell{-}1)\beta}){<}0$, and  
this is equivalent to 
$${\rm{Im}}(\frac{\alpha{+}(\ell{-}1)\beta}{\beta})
{=}{\rm{Im}}(\frac{\alpha{-}\beta}{\beta}){>}0.
$$ 
Hence, this is equivalent to ${\rm{Re}}(\beta){>}0$. 
\hfill$\square$

\bigskip

Recall that the information for the signs of the imaginary parts 
give information about the sign change of the square roots. 

Recall another factor of the determinant is given 
by the complex conjugate. 
Keeping  these in mind, we have the following.
\begin{prop}\label{on the net}
Under the $K_s$-ordered expression the assumption together with \eqref{keycond}, 
there is no singular point on lines of the 1-dimensional lattice of 
any index.
Moreover, ${:}e_*^{it\frac{1}{i\h}u_k{\ctt}v_k}{*}V_*{:}_{_K}$
is alternating $2\pi$-periodic w.r.t the variable $t{\in}{\mathbb R}$. 
In particular 
${:}e_*^{2\pi i\frac{1}{i\h}u_k{\ctt}v_k}{*}V_*{:}_{_K}{=}
{-}{:}V_*{:}_{_K}$
\end{prop}

Hence, by Proposition\,\ref{fundament} applied to the 
vertex of index 1 gives that the product via 
the connecting paths is defined to give 
$$
{:}{\e}_{00}(i_{\ell{-}1}){*}{\e}_{00}(i_{\ell}){:}_{_K}
{=}
{:}e_*^{\frac{\pi i}{i\h}
(u_{i_{\ell{-}1}}{\ctt}v_{i_{\ell{-}1}}
{+}u_{i_\ell}{\ctt}v_{i_\ell})}{:}_{_K},\,\,\text{or}\,\,
{-}{:}e_*^{\frac{\pi i}{i\h}
(u_{i_{\ell{-}1}}{\ctt}v_{i_{\ell{-}1}}
{+}u_{i_\ell}{\ctt}v_{i_\ell})}{:}_{_K}. 
$$ 
The reason of the ambiguity of $\pm$ sign is that 
the expression parameter $K$ will be so chosen that
${:}e_*^{\frac{ti}{i\h}
(u_{i_{\ell{-}1}}{\ctt}v_{i_{\ell{-}1}}
{+}u_{i_\ell}{\ctt}v_{i_\ell})}{:}_{_K}$ has a singular point 
on $t{\in}[0,\pi]$, and the $\pm$ sign is determined 
by the path avoiding the singular point.
On the other hand, since the left hand side is 
defined without ambiguity, we have the equality 
$$
{:}{\e}_{00}(i_{\ell{-}1}){*}{\e}_{00}(i_{\ell}){:}_{_K}
{=}
\gamma{:}e_*^{\frac{\pi i}{i\h}
(u_{i_{\ell{-}1}}{\ctt}v_{i_{\ell{-}1}}
{+}u_{i_\ell}{\ctt}v_{i_\ell})}{:}_{_K}, \quad  {\text{where}}\,\,
\gamma=1,\,\text{ or },\,{-1}
$$
depending on the path. By Proposition\,\ref{fundament} again, we have 
$$
{:}e_*^{\frac{is}{i\h}
u_{i_{\ell{-}2}}{\ctt}v_{i_{\ell{-}2}}}
{*}({\e}_{00}(i_{\ell{-}1}){*}{\e}_{00}(i_{\ell})){:}_{_K}
{=}
{:}e_*^{\frac{is}{i\h}
u_{i_{\ell{-}2}}{\ctt}v_{i_{\ell{-}2}}}
{*}\gamma e_*^{\frac{\pi i}{i\h}
(u_{i_{\ell{-}1}}{\ctt}v_{i_{\ell{-}1}}
{+}u_{i_\ell}{\ctt}v_{i_\ell})}
{:}_{_K}
$$
for every $s{\in}[0,\pi]$. Hence, 
at $s{=}\pi$, we have 
$$
{:}{\e}_{00}(i_{\ell{-}2})
{*}({\e}_{00}(i_{\ell{-}1}){*}{\e}_{00}(i_{\ell})){:}_{_K}
{=}
{:}\gamma e_*^{\frac{\pi i}{i\h}(u_{i_{\ell{-}2}}{\ctt}v_{i_{\ell{-}2}}{+}
u_{i_{\ell{-}1}}{\ctt}v_{i_{\ell{-}1}}
{+}u_{i_\ell}{\ctt}v_{i_\ell})}
{:}_{_K}.
$$
Repeating this procedure, we see that 
$$
{:}e_*^{\frac{\pi i}{i\h}u_{i_1}{\ctt}v_{i_1}}
{*}
\gamma e_*^{\frac{\pi i}{i\h}(u_{i_2}{\ctt}v_{i_2}{+}
\cdots{+}u_{i_{\ell}}{\ctt}v_{i_{\ell}}))}
{:}_{_K}
{=}{:}
\gamma' e_*^{\frac{\pi i}{i\h}(u_{i_1}{\ctt}v_{i_1}{+}u_{i_2}{\ctt}v_{i_2}{+}
\cdots{+}u_{i_{\ell}}{\ctt}v_{i_{\ell}}))}
{:}_{_K}
$$
and  that 
${:}e_*^{\frac{is}{i\h}u_{i_1}{\ctt}v_{i_1}}
{*}{\e}_{00}(i_2){*}\cdots{*}{\e}_{00}(i_{\ell}){:}_{_K}$
is defined and this gives at $s{=}\pi$
$$
\gamma'{:}e_*^{\frac{\pi i}{i\h}
(u_{i_1}{\ctt}v_{i_1}{+}u_{i_2}
{\ctt}v_{i_2}{+}\cdots{+}u_{i_s}{\ctt}v_{i_s}))}
{:}_{_K}.
$$ 
Hence, inductive use of 
Proposition\ref{fundament} together with Proposition\,\ref{trivrmk}
gives  
\begin{prop}\label{prod25}
Products 
${\e}_{00}(k_1){*}
{\e}_{00}(k_2){*}\cdots{*}
{\e}_{00}(k_{p})$ are welldefined in the special 
ordered expression by the path connecting products.
\end{prop}

\subsubsection{Several properties of the $K$-ordered expression of 
$e_{*}^{\frac{ti}{i\h}(u_1{\ctt}v_1{+}u_2{\ctt}v_2)}{*}V_*$.}

First, we consider the case of $V_*=1$. The $K$-ordered expression 
${:}e_{*}^{\frac{ti}{i\h}(u_1{\ctt}v_1{+}u_2{\ctt}v_2)}{:}_{_{K}}$ 
is given by computing the intertwiner 
$$
I_{_{K_0}}^{^{K}}(e^{ti}
e^{\frac{1}{i\h}\langle{\pmb u}A,{\pmb u}\rangle}){=}
\frac{e^{ti}}{\sqrt{\det{I{-}A(K{-}K_0)}}}
e^{\frac{1}{i\h}\langle{\pmb u}
\frac{1}{I{-}A(K{-}K_0)}A,{\pmb u}\rangle}.
$$
Set $\tau{=}\frac{1}{2}(e^{it}{-}1)$, $C{=}diag(\tau,\tau)$,   
$A{=}
\begin{bmatrix}
0&\tau I_2\\
\tau I_2& 0
\end{bmatrix}$, 
$K{=}
\begin{bmatrix}
S&T\\
T&S
\end{bmatrix}$, $
{}^tS{=}S,\,\,
{}^tT{=}T$ for simplicity. 

Concerning only the amplitude by \eqref{inter}, 
we only have to know $\sqrt{\det(I{-}A(K{-}K_0))}$. 
The determinant is given by elementary transformation as follows:  
\begin{equation}\label{elemtrf00}
\begin{vmatrix}
I{-}C(T{-}I)&{-}CS\\
{-}CS& I{-}C(T{-}I)
\end{vmatrix}{=}
\det\big(I{-}C(T{+}S{-}I)\big)
\det\big(I{-}C(T{-}S{-}I)\big)
\end{equation}
$$
\begin{aligned}
&{=}\tau^4
\det\big(C^{-1}{+}I{-}(T{+}S)\big)
{\det\big(C^{-1}{+}I{-}(T{-}S)\big)}\\
&{=}
\tau^4\det\Big((diag(\tau^{-1}{+}1,
\tau^{-1}{+}1)){-}(T{+}S)\Big)
\det\Big((diag(\tau^{-1}{+}1,
\tau^{-1}{+}1)){-}(T{-}S)\Big).
\end{aligned}
$$

\medskip 
Recall we have assumed that all entries of $S{+}T$ other 
than diagonal are constant $\beta$. Note at first 
by such a condition, we are considering all pairs $k,l$, 
possibly $k=l$, at the same time.
Since only $(k,l)$-submatrices of $K$ are used 
in the computation when $k,l$ are fixed, the computation 
is reduced to the case $m{=}2$. 

\medskip
For the case that the index of the vertex is $0$, that is,
$\ell{=}0$, that is the case $V_*{=}1$,  
we set  
$\tau{=}\frac{1}{2}(e^{it}{-}1)$, and we may assume 
$m{=}2$, $A{=}
\begin{bmatrix}
0&\tau I_2\\
\tau I_2& 0
\end{bmatrix}$ without loss of generality. 
The intertwiner is written as 
$$
I_{_{K_0}}^{^{K}}(e^{ti}
e^{\frac{1}{i\h}\langle{\pmb u}A,{\pmb u}\rangle}){=}
\frac{e^{ti}}{\sqrt{\det{I_2{-}A(K{-}K_0)}}}
e^{\frac{1}{i\h}\langle{\pmb u}
\frac{1}{I_2{-}A(K{-}K_0)}A,{\pmb u}\rangle}
$$
where 
$$
K{=}
\begin{bmatrix}
i\rho I{+}S'&cI{+}T'\\
cI{+}T'& i\rho I{+}S'
\end{bmatrix}, \quad 
K_0{=}
\begin{bmatrix}
0&I\\
I&0
\end{bmatrix},
$$
and 
\begin{equation}\label{65}
I{-}A(K{-}K_0){=}
\begin{bmatrix}
I{-}\tau(c{-}1)I{+}T')& {-}\tau(i\rho I{+}S')\\
{-}\tau(i\rho I{+}S')& I{-}\tau(c{-}1)I{+}T')
\end{bmatrix}.
\end{equation}

What we want to obtain is that in the $K_s$-ordered expression 
${:}e_*^{\frac{it}{i\h}
(u_k{\ctt}v_k{+}u_l{\ctt}{v_l})}{:}_{_{K_s}}$
 has singularities of order 1 on the open intervals $(0,\pi)$, and 
$(\pi,2\pi)$ for every $k,l$. 
The determinant of \eqref{65} is written 
by the elementary transformations as 
\begin{equation}\label{determinant}
\det(I{-}\tau((c{-}1{+}i\rho)I{+}T'{+}S')
\det\overline{(I{-}\bar\tau((c{-}1{+}i\rho)I{+}T'{+}S')}.
\end{equation}

Setting $S'{+}T'{=}
\begin{bmatrix}
0&\beta\\
\beta&0
\end{bmatrix}$, 
this vanishes when 
$$
(\tau^{-1}{+}1{+}\alpha)^2{-}\beta^2{=}0\quad 
{\text{or}}\quad 
(\bar\tau^{-1}{+}1{+}\alpha)^2{-}\beta^2{=}0.
$$
That is 
$$
{-}(\tau^{-1}{+}1){=}\alpha{\pm}\beta,\quad\text{or}\quad  
{-}(\bar\tau^{-1}{+}1){=}\alpha{\pm}\beta.
$$
By the condition \eqref{keycond}, we see 
${\rm{Re}}(\alpha{+}\beta)>0$. Hence,  
$$
{-}(\tau^{-1}{+}1){=}\alpha{-}\beta, \quad 
{-}(\bar\tau^{-1}{+}1){=}\alpha{-}\beta
$$
are the singular points.  Hence, we have the desired result for the case $V_*=1$. 

\bigskip
We next consider 
${:}e_*^{\frac{ti}{i\h}
(u_{k}{\ctt}v_{k}{+}u_{l}{\ctt}v_{l})}{*}V_*{:}_{_K}$ by  taking $K$ in \eqref{ordering} with 
\eqref{keycond}. 
For the case $V_{*}{\not=}1$, i.e. $\ell{\geq}1$, 
we have only to use the diagonal matrix 
$diag(\tau,{-}1,\dots,{-}1)$ instead of $\tau$ in the previous case of
$V_*=1$, where the number of ${-}1$ is the index $\ell$ 
of the vertex $V_*$. We assume that 
$V_{*}$ does not contain ${\e}_{00}(k)$ and 
${\e}_{00}(\ell)$.    
What we want to show is that  
$${:}(e_*^{\frac{ti}{i\h}
(u_k{\ctt}v_k{+}u_l{\ctt}v_l)}){*}V_{*}{:}_{_K}
$$  
is $2\pi$-periodic on ${\mathbb R}$ and it 
has singular points $\mu$, $0{<}\mu{<}\pi$, and $\nu$, 
$\pi{<}\nu{<}2\pi$.

The first factor of the 
determinant is written as  
$$
\begin{vmatrix}
\tau^{-1}{+}1{+}\alpha&\beta&\beta&\dots&\beta\\
\beta&\tau^{-1}{+}1{+}\alpha&\beta&\dots&\beta\\
\beta&\beta&\alpha&\dots&\beta\\
\vdots&\vdots&\vdots&\ddots&\vdots\\
\beta&\beta&\beta&\dots&\alpha 
\end{vmatrix}
=
\begin{matrix}
(\tau^{-1}{+}1)^2
(\alpha{+}(\ell{-}1)\beta)(\alpha{-}\beta)^{\ell{-}1}\\
\\
\quad{+}2(\tau^{-1}{+}1)
(\alpha{+}\ell\beta)(\alpha{-}\beta)^{\ell}\\
\\
\qquad{+}(\alpha{+}(\ell{+}1)\beta)(\alpha{-}\beta)^{\ell{+}1}.
\end{matrix}
$$
$$
{=}(\tau^{-1}{+}1{+}\alpha{-}\beta)
\Big((\tau^{-1}{+}1)(\alpha{+}(\ell{-}1)\beta){+}
(\alpha{+}(\ell{+}1)\beta)(\alpha{-}\beta)\Big)
(\alpha{-}\beta)^{\ell{-}1}.
$$
$$
\Big((\tau^{-1}{+}1)(\alpha{+}(\ell{-}1)\beta){+}
(\alpha{+}(\ell{+}1)\beta)(\alpha{-}\beta))\Big)
(\tau^{-1}{+}1{+}\alpha{-}\beta){=}0.
$$
It follows that the determinant vanishes if and only if 
$$
{-}(\tau^{-1}{+}1){=}
\alpha{-}\beta, \quad {\text{or}}\quad 
\frac{(\alpha{+}(\ell{+}1)\beta)(\alpha{-}\beta)}
{\alpha{+}(\ell{-}1)\beta)}
{=}\Big(1{+}\frac{2\beta}{\alpha{+}(\ell{-}1)\beta}\Big)
(\alpha{-}\beta).
$$
Since $\alpha{-}\beta{=}id_+$, $d_+{>}0$, the same argument as 
in the proof of Lemma\,\ref{estimate1} gives that  
$${\rm{Re}}\,
\Big(1{+}\frac{2\beta}{\alpha{+}(\ell{-}1)\beta}\Big)
(\alpha{-}\beta) >0.
$$ 
Hence, we have only to assume \eqref{keycond}.

\medskip
For the second factor, the determinant is given by the 
complex conjugate. Hence, the requested conditions are satisfied by 
\eqref{keycond}.
The singular points together with the second factor are given by 
$$
e^{is}{=}\frac{id_+{+}1}{id_+{-}1},\quad 
\frac{id_+{-}1}{id_+{+}1}. 
$$

\begin{prop}\label{singpoints}
Under the condition 
\eqref{keycond} for $K$,  
${:}e_*^{\frac{ti}{i\h}
(u_k{\ctt}v_k{+}u_l{\ctt}v_l)}{*}V_{*}{:}_{_K}$, $k{\not=}l$, 
has singular points at $t_0{\in}(0,\pi)$ and $2\pi{-}t_0$ 
for every vertex $V_{*}$, and 
$${:}e_*^{\frac{2\pi i}{i\h}
(u_k{\ctt}v_k{+}u_l{\ctt}v_l)}{*}V_{*}{:}_{_K}{=}
{:}V_{*}{:}_{_K}
$$  
for every $1\leq k,\,l\leq m$, if we take the anit-clockwise 
half-circle path avoiding the singular point.
\end{prop}

The $\pm$ sign of ${:}e_*^{\frac{ti}{i\h}
(u_{k}{\ctt}v_{k}{+}u_{l}{\ctt}v_{l})}{*}V_*{:}_{_K}$ is 
determined by the path avoiding the singularity.

\subsubsection{Determinant equation of two variables}

Here, we show that 
${:}e_*^{\frac{i}{i\h}
(t_1u_k{\ctt}v_k{+}t_2u_l{\ctt}v_l)}{*}V_{*}{:}_{_K}$, $k{\not=}l$, 
has no singular point other than the singular point lying in the 
diagonal $(t_0,t_0){\in}[0,\pi]\times[0,\pi]$. Here, we assume that 
$V_{*}$ does not contain ${\e}_{00}(k)$ and 
${\e}_{00}(\ell)$.   

As singular points are given by zeros of \eqref{elemtrf}, 
in this section we consider the equation  
\begin{equation}\label{master00}
\begin{vmatrix}
\tau_1^{-1}{+}1{+}\alpha&\beta&\beta&\dots&\beta\\
\beta&\tau_2^{-1}{+}1{+}\alpha&\beta&\dots&\beta\\
\beta&\beta&\alpha&\dots&\beta\\
\vdots&\vdots&\vdots&\ddots&\vdots\\
\beta&\beta&\beta&\dots&\alpha 
\end{vmatrix}
\begin{vmatrix}
\tau_1^{-1}{+}1{+}\bar\alpha&\bar\beta&\bar\beta&\dots&\bar\beta\\
\bar\beta&\tau_2^{-1}{+}1{+}\bar\alpha&\bar\beta&\dots&\bar\beta\\
\bar\beta&\bar\beta&\bar\alpha&\dots&\bar\beta\\
\vdots&\vdots&\vdots&\ddots&\vdots\\
\bar\beta&\bar\beta&\bar\beta&\dots&\bar\alpha 
\end{vmatrix}
{=}0
\end{equation}
of two variables under the restriction $\tau_i\in i{\mathbb R}$. 
For simplicity, we set 
$ix{=}\tau_1^{-1}{+}1$, $iy{=}\tau_2^{-1}{+}1$.
The first factor is written as  
$$
-xy(\alpha{+}(\ell{-}1)\beta)(\alpha{-}\beta)^{\ell{-}1}
{+}i(x{+}y)
(\alpha{+}\ell\beta)(\alpha{-}\beta)^{\ell}
{+}(\alpha{+}(\ell{+}1)\beta)(\alpha{-}\beta)^{\ell{+}1}{=}0, 
\,\,\,(\ell\geq 0). \quad i.e.
$$
$$
-xy
{+}i(x{+}y)
\frac{(\alpha{+}\ell\beta)(\alpha{-}\beta)}
{(\alpha{+}(\ell{-}1)\beta)}
{+}\frac{(\alpha{+}(\ell{+}1)\beta)(\alpha{-}\beta)^2}
{(\alpha{+}(\ell{-}1)\beta)}{=}0. 
$$
If we set 
$A{=}\frac{(\alpha{+}\ell\beta)(\alpha{-}\beta)}
{(\alpha{+}(\ell{-}1)\beta)}$ and  
$$
B^2{=}\Big(\frac{(\alpha{+}\ell\beta)(\alpha{-}\beta)}
{(\alpha{+}(\ell{-}1)\beta)}\Big)^2
{-}\frac{(\alpha{+}(\ell{+}1)\beta)(\alpha{-}\beta)^2}
{(\alpha{+}(\ell{-}1)\beta)}{=}
\Big(\frac{(\alpha{-}\beta)\beta}{\alpha{+}(\ell{-}1)\beta}\Big)^2, 
$$ 
then the first factor is rewritten as 
$(ix{+}A)(iy{+}A){-}B^2$. Similarly,  
the second factor is written as  
$(ix{+}\bar A)(iy{+}\bar A){-}\bar B^2$.

\bigskip
We put in \eqref{keycond}  
the condition that $\alpha{-}\beta$ is a positive pure imaginary 
number, and set $\alpha{-}\beta{=}id_+$. 
Setting $A{=}A_0{+}iA_1, \,\, B{=}B_0{+}iB_1$, we have 
$$
A_0{=}{-}d_+{\rm{Im}}\frac{\beta}{\alpha{+}(\ell{-}1)\beta},\quad
A_1{=}d_+(1{+}{\rm{Re}}\frac{\beta}{\alpha{+}(\ell{-}1)\beta}),
$$

$$
B_0={-}d_+{\rm{Im}}\frac{\beta}{\alpha{+}(\ell{-}1)\beta},\quad
B_1{=}d_+{\rm{Re}}\frac{\beta}{\alpha{+}(\ell{-}1)\beta}.
$$
Hence, we see $A_0{=}B_0$, $A_1{-}B_1{=}d_+$. 
Let $\alpha{=}a{+}i\rho, \beta{=}a{+}ib$.

Note now that 
\begin{equation*}
\begin{aligned}
&{\rm{Im}}\frac{\beta}{\alpha{+}(\ell{-}1)\beta}{<}0 
\Longleftrightarrow 
{\rm{Im}}\frac{\alpha}{\beta}{>}0
\Longleftrightarrow a(\rho{-}b){>}0,\\
&{\rm{Re}}\frac{\beta}{\alpha{+}(\ell{-}1)\beta}{>}0 
\Longleftrightarrow 
{\rm{Re}}(\ell{-}1{+}\frac{\alpha}{\beta}){>}0
\Longleftrightarrow \ell{-}1{+}\frac{a^2{+}\rho b}{a^2{+}b^2}{>}0. 
\end{aligned}
\end{equation*}
Since $\rho b{>}0$ is assumed, if $\ell{\geq}1$ then
$\ell{-}1{+}\frac{a^2{+}\rho b}{a^2{+}b^2}{>}0$. 
For $\ell{=}0$, if $\rho{-}b{>}0$ then 
${-}1{+}\frac{a^2{+}\rho b}{a^2{+}b^2}{>}0$.

In addition, suppose in addition that 
$\alpha{=}a{+}i\rho=c{+}i\rho, \,\,\beta{=}a{+}ib$ 
satisfy that $a{>}b{>}0$, $\rho{>}b{>}0$. 
Then we easily see that $A_0{=}B_0{>}0$,  
$A_1^2{>}B_1^2$ for every $\ell{\geq}0$. Note that 
all additional conditions other than the condition $c{=}a$ 
are open conditions, hence we have three real dimensions 
of freedom.

\bigskip

Suppose we have an equation of pure imaginary variables $ix,iy$  
\begin{equation}\label{mastereq}
(ix{+}A)(iy{+}A){-}B^2{=}0, 
\quad 
A, B{\in}{\mathbb C}, \quad x, y\in{\mathbb R}.
\end{equation}
Set $A=A_0{+}iA_1$, $A=B_0{+}iB_1$

\begin{lem}\label{keykey}
If $A_0=B_0\not=0$, then 
taking the imaginary part and  the real part of the equation \eqref{mastereq}, we have 
$$
x{+}y=2(B_1{-}A_1),\quad xy=(B_1{-}A_1)^2.
$$
This shows that the solution of \eqref{mastereq} is degenerate.
\end{lem}

It follows that  
${:}e_*^{\frac{i}{i\h}
(t_1u_k{\ctt}v_k{+}t_2u_l{\ctt}v_l)}{*}V_{*}{:}_{_K}$, $k{\not=}l$, 
has no other singular point than those sitting in the diagonal set 
$(t_0,t_0){\in}[0,\pi]\times[0,\pi]$.

\bigskip

\subsubsection{Emergence of Clifford algebra}

In the case $V_*$ does not contain ${\e}_{00}(k)$ but it contains  
${\e}_{00}(\ell)$, we see by Proposition\,\ref{on the net} that ${\e}_{00}(\ell){*}V_*={-}V'_*$. Hence, 
$V_*={\e}_{00}(\ell){*}V_*'$ where $V'_*$ does not contain 
${\e}_{00}(k),\,\,{\e}_{00}(\ell)$. Thus, by
Proposition\,\ref{singpoints} and \eqref{exchange} we see 
$$
{\e}_{00}(k){*}{\e}_{00}(\ell){*}V_*=
{\e}_{00}(k){*}{\e}_{00}(\ell){*}{\e}_{00}(\ell){*}V_*'=-{\e}_{00}(k){*}V'_*.
$$
On the other hand 
$$
{\e}_{00}(\ell){*}{\e}_{00}(k){*}V_*=
{\e}_{00}(\ell){*}{\e}_{00}(k){*}{\e}_{00}(\ell){*}V_*'=
{-}{\e}_{00}(\ell){*}{\e}_{00}(\ell){*}{\e}_{00}(k){*}V_*'={\e}_{00}(k){*}V_*'
$$

\medskip
In the case $V_*$ contains ${\e}_{00}(k)$,\,\, 
${\e}_{00}(\ell)$, we set ${\e}_{00}(k){*}{\e}_{00}(\ell){*}V_*=V'_*$  
where $V'_*$ does not contain 
${\e}_{00}(k),\,\,{\e}_{00}(\ell)$. In this situation, we have 
$$
{\e}_{00}(k){*}{\e}_{00}(\ell){*}V'_*={-}{\e}_{00}(\ell){*}{\e}_{00}(k){*}V'_*.
$$
Suppose  
$V_*={\e}_{00}(k){*}{\e}_{00}(\ell){*}V'_*$, then  
$$
\begin{aligned}
{\e}_{00}(k){*}{\e}_{00}(\ell){*}V_*&=
{\e}_{00}(k){*}{\e}_{00}(\ell){*}{\e}_{00}(k){*}{\e}_{00}(\ell){*}V'_*\\
&=-{\e}_{00}(k){*}{\e}_{00}(\ell){*}{\e}_{00}(\ell){*}{\e}_{00}(k){*}V'_*
={\e}_{00}(k){*}{\e}_{00}(k){*}V'_*{=}{-}V'_*
\end{aligned}
$$
On the other hand, 
$$
{\e}_{00}(\ell){*}{\e}_{00}(k){*}V_*=
{\e}_{00}(\ell){*}{\e}_{00}(k){*}{\e}_{00}(k){*}{\e}_{00}(\ell){*}V'_*
=-{\e}_{00}(\ell){*}{\e}_{00}(\ell){*}V'_*{=}V'_*
$$

Thus, Theorem\,\ref{EXPspecial} is proved 
by these observation.  Namely, we see   

\begin{thm}\label{cliffmmm}
In a $K_s$-ordered expression, 
$(H\!ol({\mathbb C}^{2m}),{*_{_{K_s}}})$  contains the
Clifford algebra $Clif\!f(m)$. 
\end{thm}

Since $({\e}_{00}(k){*}{\e}_{00}(\ell))_*^2=-1={\e}_{00}(k)_*^2$ in
$K_s$-ordered expression, we see in particular 
\begin{cor}\label{cliffmmm}
In the $K_s$-ordered expression, 
the group ${\mathfrak P}^{(4)}_{_{K_s}}$ coisncides 
with ${\mathfrak P}^{(2)}_{_{K_s}}$. It follows that 
${\mathfrak P}^{(2)}_{_{K_s}}$ is a connected double cover of 
$SO(m,{\mathbb C})$, hence 
${\mathfrak P}^{(2)}_{_{K_s}}\cong Spin(m){\otimes}{\mathbb C}$.
\end{cor}

\end{document}